\documentclass[aps,prl,amsmath,amssymb,twocolumn,floatfix,superscriptaddress]{revtex4-1}
\usepackage{graphicx}
\usepackage{bm}
\usepackage{latexsym}
\usepackage{epsf}
\usepackage{rotating}
\usepackage{epsfig,graphics,rotate,color}
\usepackage{wrapfig}
\usepackage{gensymb}
\usepackage{amssymb}
\usepackage{amsmath}
\usepackage{amsfonts}
\usepackage{array,hhline,dcolumn}%
\usepackage[normalem]{ulem}
\usepackage{color}
\usepackage{color}
\usepackage{hyperref}
\usepackage{graphicx}
\usepackage{placeins}
\usepackage{mwe}
\usepackage{lineno}
\begin{document}

\title{First Measurement of Electron Neutrino Scattering Cross Section on Argon}

\date{\today}

\author{R.~Acciarri}
\affiliation{Fermi National Accelerator Lab, Batavia, Illinois 60510, USA}

\author{C.~Adams}
\affiliation{Argonne National Lab, Lemont, Illinois 60439, USA}

\author{J.~Asaadi}
\affiliation{University of Texas at Arlington, Arlington, Texas 76019, USA}

\author{B.~Baller}
\affiliation{Fermi National Accelerator Lab, Batavia, Illinois 60510, USA}

\author{V.~Basque}
\affiliation{University of Manchester, Manchester M13 9PL, United Kingdom}

\author{T.~Bolton}
\affiliation{Kansas State University, Manhattan, Kansas 66506, USA}

\author{C.~Bromberg}
\affiliation{Michigan State University, East Lansing, Michigan 48824, USA}

\author{F.~Cavanna}
\affiliation{Fermi National Accelerator Lab, Batavia, Illinois 60510, USA}

\author{D.~Edmunds}
\affiliation{Michigan State University, East Lansing, Michigan 48824, USA}

\author{R.S.~Fitzpatrick}
\email{roryfitz@umich.edu}
\affiliation{University of Michigan, Ann Arbor, Michigan 48109, USA}

\author{B.~Fleming}
\affiliation{Yale University, New Haven, Connecticut 06520, USA}

\author{P.~Green}
\affiliation{University of Manchester, Manchester M13 9PL, United Kingdom}

\author{C.~James}
\affiliation{Fermi National Accelerator Lab, Batavia, Illinois 60510, USA}

\author{I.~Lepetic}
\affiliation{Illinois Institute of Technology, Chicago, Illinois 60616, USA}
\affiliation{Rutgers University, Piscataway New Jersey 08854, USA}

\author{B.R.~Littlejohn}
\affiliation{Illinois Institute of Technology, Chicago, Illinois 60616, USA}

\author{X.~Luo}
\affiliation{University of California, Santa Barbara, California, 93106, USA}

\author{O.~Palamara}
\affiliation{Fermi National Accelerator Lab, Batavia, Illinois 60510, USA}

\author{G.~Scanavini}
\affiliation{Yale University, New Haven, Connecticut 06520, USA}

\author{M.~Soderberg}
\affiliation{Syracuse University, Syracuse, New York 13244, USA}

\author{J.~Spitz}
\affiliation{University of Michigan, Ann Arbor, Michigan 48109, USA}

\author{A.M.~Szelc}
\affiliation{University of Manchester, Manchester M13 9PL, United Kingdom}

\author{W.~Wu}
\affiliation{Fermi National Accelerator Lab, Batavia, Illinois 60510, USA}

\author{T.~Yang}
\affiliation{Fermi National Accelerator Lab, Batavia, Illinois 60510, USA}

\collaboration{The ArgoNeuT Collaboration}
\noaffiliation

\begin{abstract}

We report the first electron neutrino cross section measurements on argon, based on data collected by the ArgoNeuT experiment running in the GeV-scale NuMI beamline at Fermilab. A flux-averaged $\nu_e + \overline{\nu}_e$ total and a lepton angle differential cross section are extracted using 13~$\nu_e$ and $\overline{\nu}_e$ events identified with fully-automated selection and reconstruction. We employ electromagnetic-induced shower characterization and analysis tools developed to identify $\nu_e/\overline{\nu}_e$-like events among complex interaction topologies present in ArgoNeuT data ($\langle E_{\bar{\nu}_e} \rangle = 4.3$ GeV and $\langle E_{\nu_e} \rangle = 10.5$ GeV). The techniques are widely applicable to searches for electron-flavor appearance at short- and long-baseline using liquid argon time projection chamber technology. Notably, the data-driven studies of GeV-scale $\nu_e/\overline{\nu}_e$ interactions presented in this Letter probe an energy regime relevant for future DUNE oscillation physics.

\end{abstract}
\maketitle
While neutrino mass and mixing has enjoyed a bounty of rich discoveries over the past few decades, a number of questions remain. Most notably, the ordering of the neutrino mass states, the value of the CP-violating phase ($\delta_{CP}$), and the possibility of new degrees of freedom driving oscillations (e.g. $\nu_{e,\mu,\tau}\rightarrow \nu_s$), remain open questions. Electron neutrino identification and characterization is essential to the $\nu_\mu \rightarrow \nu_e$ and $\overline{\nu}_\mu \rightarrow \overline{\nu}_e$ appearance-based short- and long-baseline experiments seeking answers to these questions \cite{sbn_prop, dune_tdr, nova_osc, t2k_osc, hkk_osc}. The precision required for these measurements calls for high-resolution detection techniques like the liquid argon time projection chamber (LArTPC) technology deployed by many current and upcoming experiments. In particular, the SBN Program at Fermilab~\cite{sbn_prop} studies the possibility of a sterile flavor participating in oscillations, and DUNE~\cite{dune_tdr} seeks to determine the neutrino mass hierarchy and extract $\delta_{CP}$, both using LArTPCs. 

Exploring these physics topics with LArTPC experiments requires careful reconstruction of $\nu_e$ and $\overline{\nu}_e$ interactions, often difficult to identify with automated methods. Even with LArTPC technology and its ability to provide mm-scale-resolution pictures of the events in question, hit and cluster finding, shower formation, and finally, neutrino energy reconstruction and flavor identification, remain challenging. Algorithms for effectively interpreting the abundance of information provided in LArTPC data are critical for extracting physics results. This is particularly true for DUNE, which will rely on the inclusive set of all $\nu_e$/$\overline{\nu}_e$ charged current (CC) interactions in the few-GeV energy range~\cite{dune_tdr} rather than an exclusive CC quasi-elastic-like signal channel. The selection must accommodate substantial contributions from the varying event topologies associated with quasi-elastic, resonant, and deep inelastic scattering, and significant effects from nuclear physics, including multi-nucleon correlations and final state interactions~\cite{pdg_2018}. Background events in a $\nu_e$/$\overline{\nu}_e$ search in DUNE also present challenges; even for underground detectors with low or negligible cosmic contamination, the electromagnetic (EM) showers characteristic of $\nu_e$/$\overline{\nu}_e$ events are readily mimicked by numerous neutrino-induced background processes, especially $\nu_\mu/\overline{\nu}_\mu$ CC and neutral current (NC) interactions featuring $\pi^0$ $\rightarrow \gamma\gamma$ and non-negligible $\Delta$ $\rightarrow \mathrm{N}\gamma$ contributions. These energy reconstruction and background issues directly affect oscillation measurements. For example, while DUNE is expected to be statistics-limited early on with exposures less than 100~kt$\cdot$MW$\cdot$year, energy-scale, flux, and interaction model systematic uncertainties will quickly take the lead in the $\delta_{CP}$ measurement uncertainty budget~\cite{dune_tdr}. 

On the way to efficient electron-flavor reconstruction with minimal background in LArTPC neutrino experiments, data-based studies at the GeV-scale are largely absent. This Letter is the first to report $\nu_e$/$\overline{\nu}_e$ measurements extracted from GeV-scale neutrino beam data using automated methods.

Previously, ArgoNeuT demonstrated that topological information alone could be used to identify electron neutrino candidates by rejecting gamma backgrounds based on the characteristic gap expected between the neutrino interaction vertex and the beginning of a gamma-induced shower due to the large (relative to LArTPC spatial resolution) conversion length of 18~cm in liquid argon~\cite{argoneut_nue}. It was further shown, using samples of events selected by visual scanning methods containing either an electron or gamma candidate, that vertex $dE/dx$ could be used to separate electrons from gammas, a notable milestone in LArTPC reconstruction for exploiting the wealth of charge and spatial detail provided by the technology. However, these strategies are quickly complicated by interactions with high multiplicity where hadronic overlap with EM showers can obscure the essential gap and $dE/dx$ information close to the vertex.

Towards the total $\nu_e + \overline{\nu}_e$ CC cross section reported in the latter part of this Letter, we first provide a short description of the ArgoNeuT detector and a detailed explanation of the EM shower reconstruction, background and systematics estimation, and signal extraction procedures employed, providing an analysis framework for future LArTPC-based $\nu_e/\overline{\nu}_e$ appearance searches. While previous $\nu_e/\overline{\nu}_e$ CC studies in LArTPCs have relied only on topological and calorimetric information specific to the neutrino interaction vertex \cite{argoneut_nue}, this study broadens the scope of classification tools to take advantage of the entire EM shower topology, a necessary step toward developing inclusive $\nu_e/\overline{\nu}_e$ CC selection strategies for GeV-scale neutrino interactions in the presence of significant background.



The ArgoNeuT LArTPC experiment at Fermilab collected data in the NuMI beamline just upstream of the MINOS near detector~\cite{minos_detector} in 2009-2010, with the vast majority ($1.25\times 10^{20}$~POT) taken in ``low-energy antineutrino mode" ($\langle E_{\bar{\nu}_e} \rangle = 4.3$ GeV with 68\% falling between 1.0 and 6.5 GeV and $\langle E_{\nu_e} \rangle = 10.5$ GeV with 68\% falling between 2.5 and 21.5 GeV)~\cite{numi_flux2}. ArgoNeuT featured a $40\times 47 \times 90$~cm$^3$ [vertical, drift, horizontal (beam)] TPC at 481~V/cm with 240 induction and 240 collection wires separated by 4~mm and sampled at a rate of 5 MHz by the readout electronics. The detector is described in detail in Ref.~\cite{argoneut_detector}. 


Neutrino interactions in ArgoNeuT are simulated using the GENIE~\cite{genie} neutrino event generator in combination with \textsc{GEANT}4-based~\cite{geant} detector and particle propagation models. Neutrino and antineutrino fluxes from the NuMI beam are provided by the MINER$\nu$A Collaboration~\cite{numi_flux2}. After event simulation, interactions in the ArgoNeuT detector are first reconstructed using the LArSoft software package~\cite{larsoft}. The algorithms, described in detail in Ref.~\cite{cc1pi}, proceed in the following steps: 1) noise removal and deconvolution of raw wire signals to correct for electronics and field response, 2) hit finding, 3) clustering of hits on each plane based on proximity to one another, 4) reconstruction of three-dimensional (3D) tracks by matching clusters across wire planes with temporal consistency, and 5) calorimetric reconstruction. 

Custom reconstruction tools then use the output of the standard software package to build candidate EM showers. The shower reconstruction algorithm relies on two objects produced by the standard tools: 1) 3D tracks with associated vertex and direction information, and 2) clusters of hits on each plane tagged as ``shower-like" or ``track-like" based on the measure of multiple coulomb scattering along the clustered hits, the size of the cluster, and its proximity to other clusters.

The shower reconstruction algorithm used in this analysis is designed to reconstruct electrons in ArgoNeuT. While the subsequently described selection procedure could be applied to any population of electron neutrino candidate events, we focus on reconstructing only the leading shower in each neutrino interaction. Particularly, we make use of the reliable 3D track reconstruction for defining the vertex and direction of a shower. The shower reconstruction builds candidate electron showers around reconstructed 3D tracks by looking for shower-like clusters of hits in close proximity to the track axis; the hit and proximity thresholds for finding candidate showers were optimized to maximize reconstruction completeness, purity, and efficiency for electrons specifically.

After reconstruction, a set of quality cuts is applied to the data. First, a filter is applied to reject events with a muon reconstructed in the downstream MINOS near detector that when projected backward crosses the ArgoNeuT detector active volume, indicative of a background $\nu_\mu$/$\overline{\nu}_\mu$ CC interaction. The vertex of each candidate signal electron must lie within the fiducial volume defined to be 3 cm inside the anode and cathode planes, 4 cm from the top and bottom boundaries of the TPC, 6 cm from the upstream face of the detector, and 20 cm from the downstream face of the detector. The vertex must be at least 20 cm from the back of the detector to give candidate electrons enough space to begin exhibiting shower-like qualities, motivated by the radiation length ($X_0$=14~cm) in argon. Additionally, reconstructed showers must have $\cos(\theta_z) > 0.05$, resulting in a negligible loss of phase space in favor of removing backward-going failures of reconstruction more prominent in data due to electronics noise. To remove events with through-going muons  from neutrino interactions upstream of the TPC, we reject any event with a reconstructed 3D cluster that falls within 2.5~cm of the upstream face. Finally, we require that the closest hit in a candidate shower on each plane be within 2~cm of the reconstructed vertex to remove track reconstruction failures.

\begin{figure}[tbp]
\centering 
\includegraphics[width=.45\textwidth,trim={25 0 0 25}]{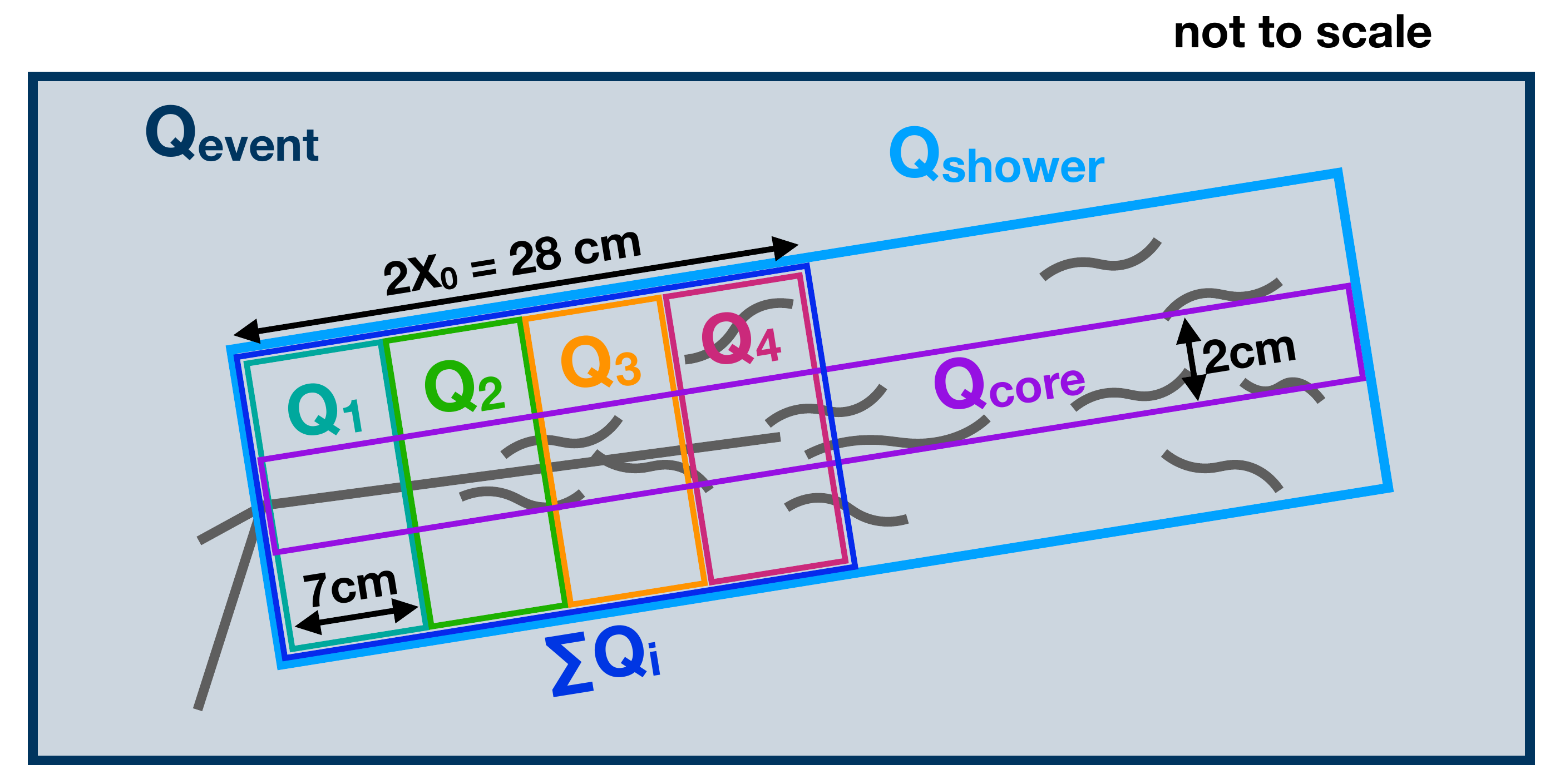}
\caption{\label{fig:shower}The topological charge regions defined for each reconstructed shower, intended to characterize the transverse and longitudinal development of the candidate shower. The cells defined $Q_1$ through $Q_4$ extend in the transverse direction to include all hits identified as part of the candidate shower. $Q_\mathrm{core}$ extends to include all shower hits in the longitudinal direction. $Q_\mathrm{shower}$ extends in both directions to include all hits identified as part of the candidate shower. The distances defined are in two dimensions.}
\end{figure}

\begin{figure}[tbp]
\centering 
\includegraphics[width=.45\textwidth,trim={20 10 10 20}]{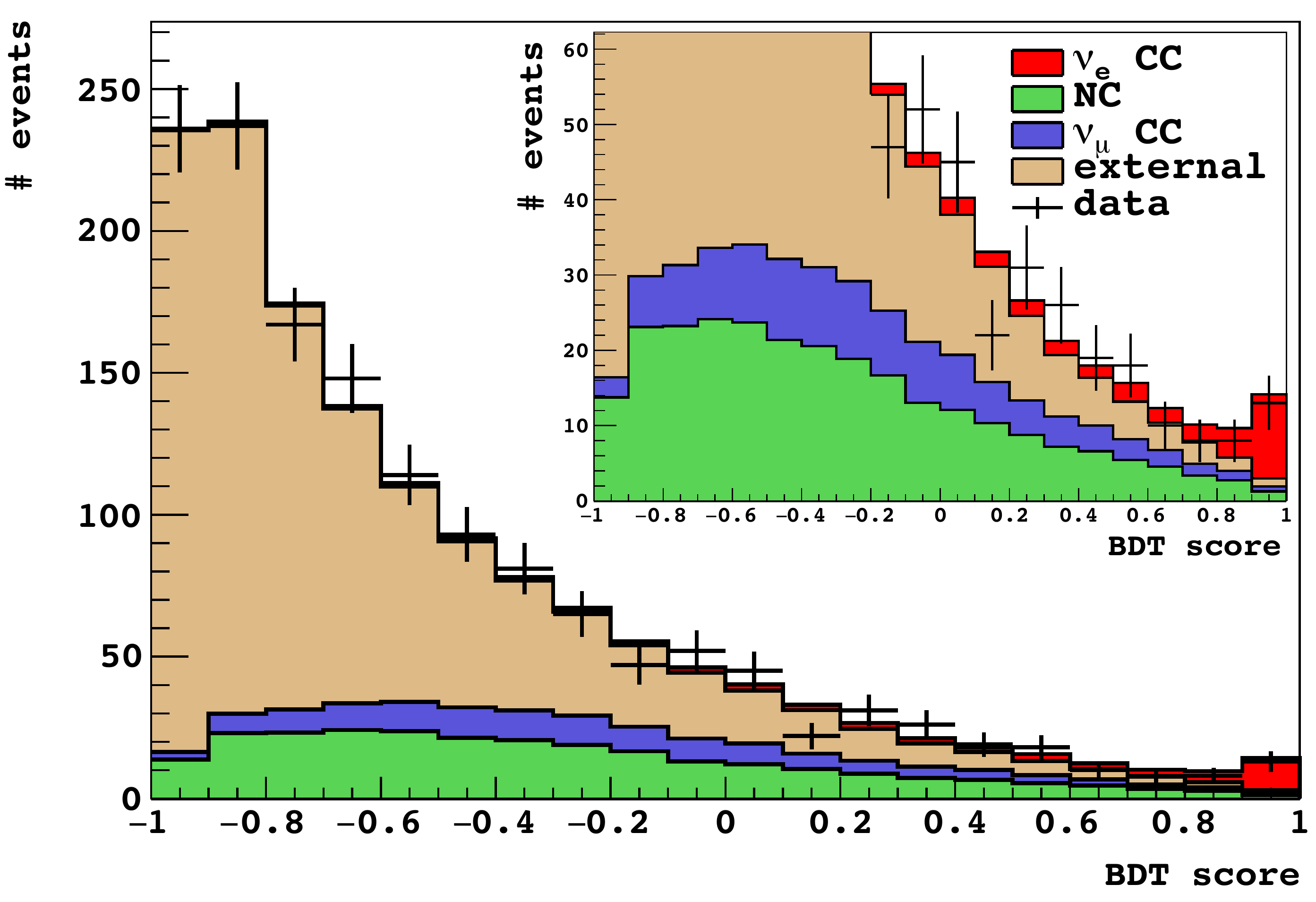}
\caption{\label{fig:bdt} The distribution of BDT scores for data and simulation. The signal selection in this analysis uses events with BDT score $> 0.9$. The inset shows the same information zoomed in to better show this signal region.}
\end{figure}

The calorimetric discrimination techniques described here could be applied to any reconstructed shower object, independent of the reconstruction algorithm. We simply assume a reconstructed shower possesses 1) a collection of hits on at least one plane, 2) a vertex, and 3) a direction.  In the small ArgoNeuT detector, the rarity of complete shower containment prohibits the use of total charge (e.g., for complete shower characterization and energy reconstruction). We instead use charge ratios constructed from topological regions of charge, shown in Figure \ref{fig:shower}, to characterize the shape and evolution of each candidate electron shower \cite{pdg_2018}. For example, longitudinal development of the shower is modeled by defining the ratios $Q_n/ \sum_i Q_i$ where $n = 1, 2, 3, 4$, and transverse shower development is characterized using $Q_\mathrm{core}/Q_\mathrm{shower}$. These topologically motivated charge ratios are powerful discriminators, along with vertex $dE/dx$, for selecting signal $\nu_e/\overline{\nu}_e$ CC events among backgrounds involving a variety of event classes. These include difficult-to-reconstruct background deep-inelastic events often characterized by multiple overlapping tracks and EM activity. The following variables are defined for $\nu_e/\overline{\nu}_e$ CC classification using a boosted decision tree (BDT): three angles [$\cos(\theta_x), \cos(\theta_y),$ and $\cos(\theta_z)$], $Q_\mathrm{shower}/Q_\mathrm{event}$, $\sum_i Q_i/Q_\mathrm{shower}$, $Q_\mathrm{core}/Q_\mathrm{shower}$, $Q_n/\sum_i Q_i$, and vertex $dE/dx$, calculated by taking the median charge in the first 4 cm of the track~\cite{argoneut_nue}. All charge variables are defined using the collection plane only. The output of the BDT trained using these quantities is shown in Figure \ref{fig:bdt}. The three most important inputs for separating signal and background, all with approximately equal impact, are $Q_\mathrm{shower}/Q_\mathrm{event}$, $Q_\mathrm{core}/Q_\mathrm{shower}$ and vertex $dE/dx$. The distance between the neutrino vertex and EM shower start is not used in this analysis for signal identification; the high neutrino energies and resulting large track multiplicities complicate automated gap reconstruction, yielding weak separation power between electrons and gammas.

One difficulty in this analysis is the small size of the ArgoNeuT  detector. We find that a significant background comes from EM-like activity in the detector produced by interactions originating outside of the detector active volume. This is a complication unique to ArgoNeuT, where it is impossible to move sufficiently far from the edge of the detector to reject a significant fraction of these outside backgrounds while simultaneously maintaining satisfactory signal statistics. Additionally, we find that the external background is underestimated in the ArgoNeuT simulation, which only generates neutrino interactions that occur with and inside the cryostat. While the simulation reproduces the energy and topological characteristics of external EM-like backgrounds in the detector, it misrepresents the total quantity of these backgrounds. To correct for this deficit and constrain the external background contribution in the strict $\nu_e/\overline{\nu}_e$ selection region, the external background is scaled as a linear function of BDT score, derived using a data-simulation comparison sideband with score $< 0$. The data-driven function is motivated by the fact that external backgrounds tend to look topologically distinct from signal, a characteristic which is quantitatively described by decreasing BDT score, a proxy for event topology. 

To reduce the impact of the uncertainty associated with the background scaling on the final selection, we have limited our signal definition to events with BDT score $> 0.9$, a bin with low external and total background that yields the most significant signal selection. A conservative 100\% uncertainty on the quantity of external background is included in the systematic error, and is also consistent with the errors on the fitted constraint when extrapolated to the signal region. Other systematic uncertainties considered include those associated with the neutrino interaction model, found by varying a set of relevant parameters in GENIE independently according to Ref.~\cite{genie}, in addition to uncertainties in the integrated flux, collected POT, and number of target argon nuclei. Given the low statistics of our measurement, statistical uncertainties dominate the results reported here. 

\begin{figure}[tbp]
\centering 
\includegraphics[width=.45\textwidth,trim={20 10 0 10}]{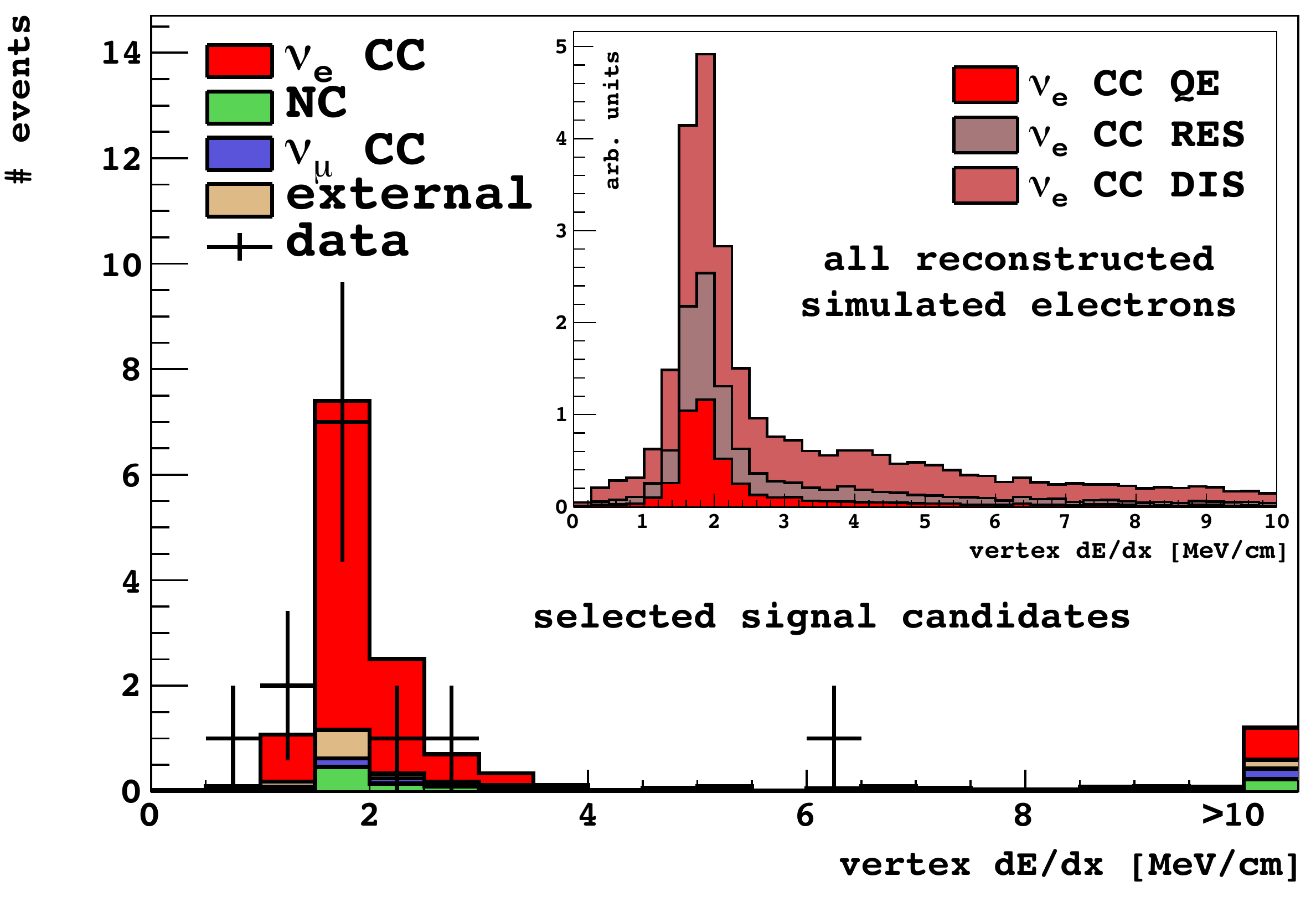}
\caption{\label{fig:dedx} Reconstructed vertex $dE/dx$ for data and simulation after selection. The inset shows the vertex $dE/dx$ distribution for electrons reconstructed from a sample of simulated $\nu_e/\overline{\nu}_e$ events broken up by interaction mechanism, demonstrating that the vertex $dE/dx$ tail is mainly from deep inelastic scattering.}
\end{figure}

\begin{figure}[tbp]
\centering 
\includegraphics[width=.49\textwidth]{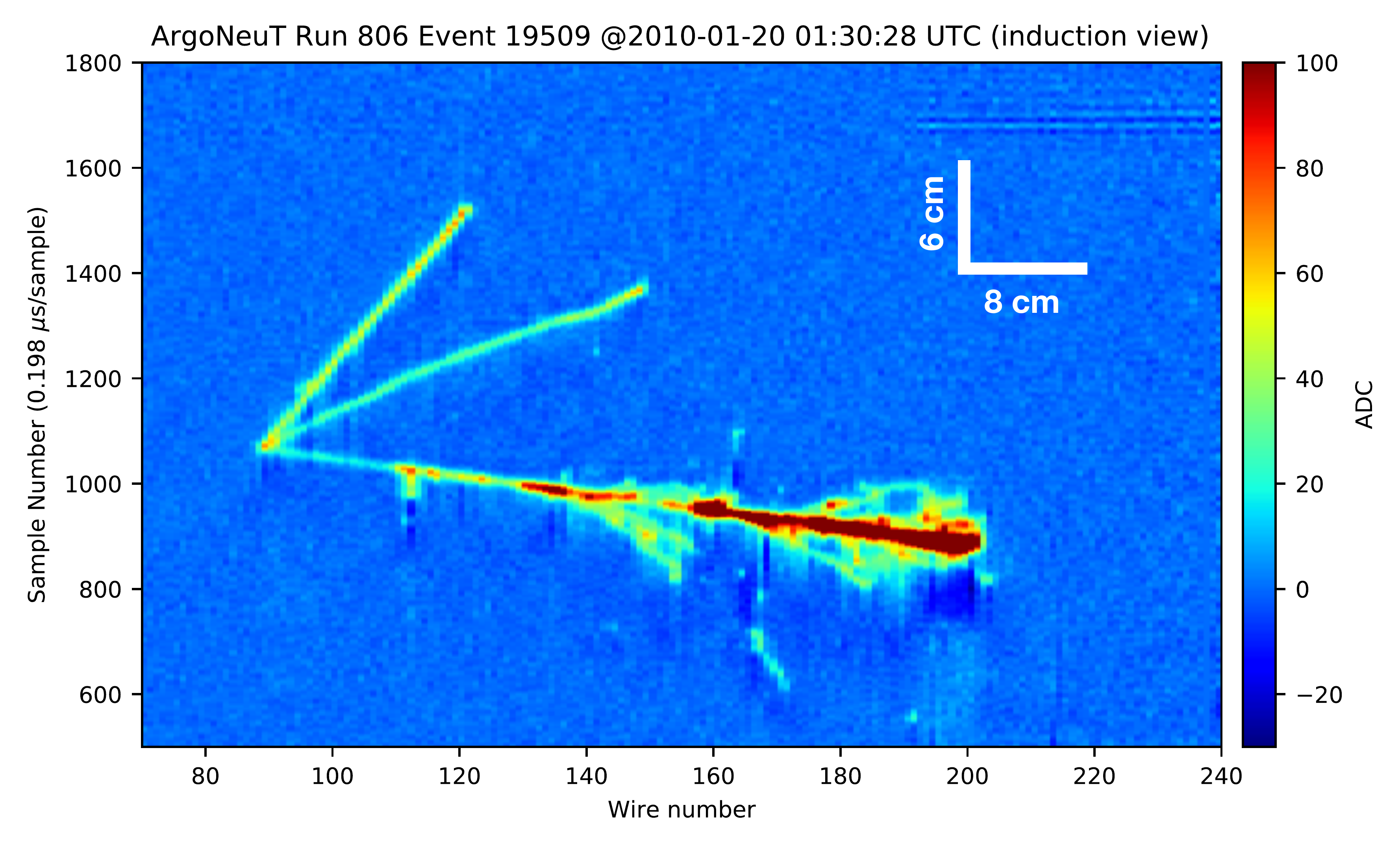}
\includegraphics[width=.49\textwidth]{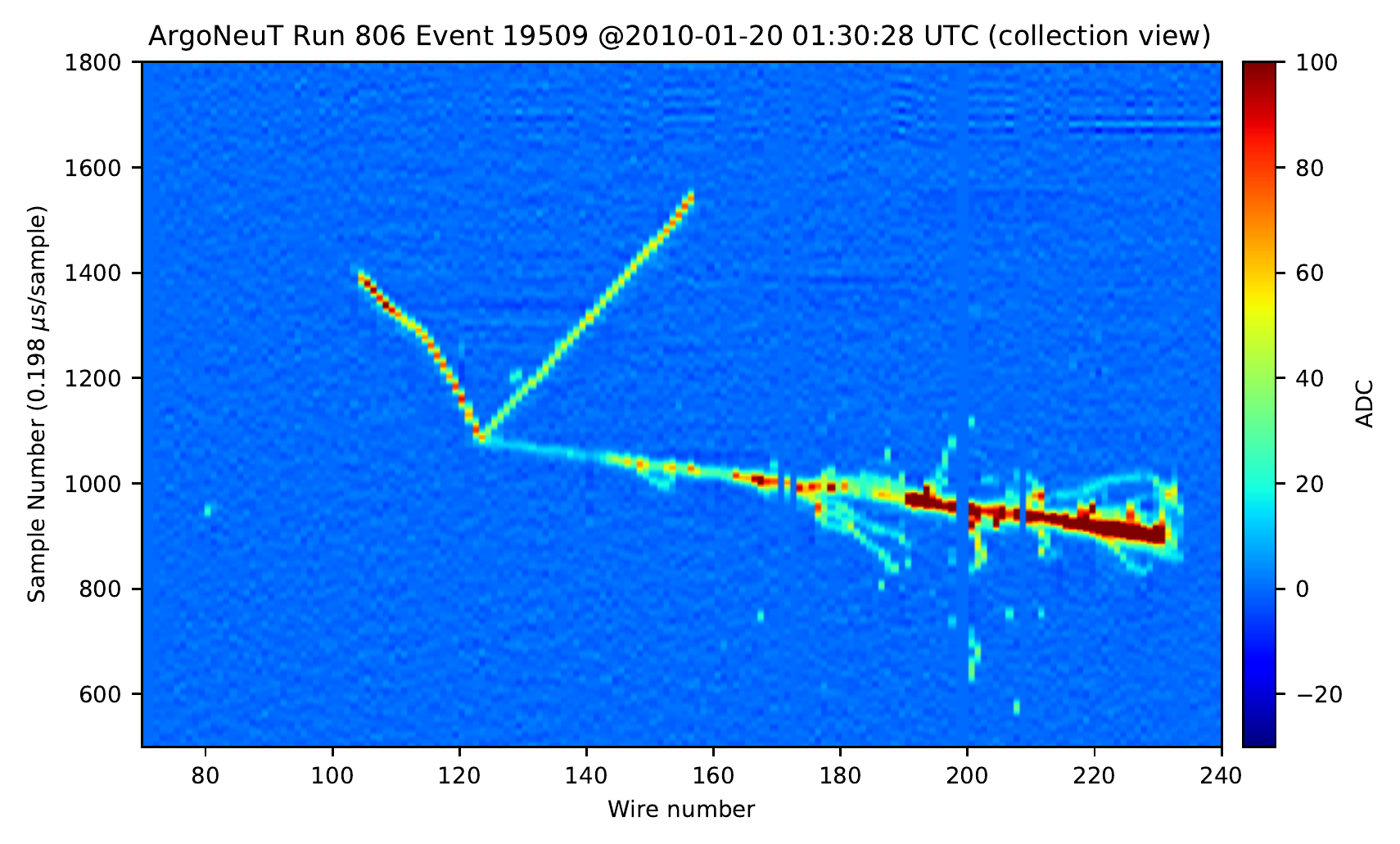}
\caption{\label{fig:nue} A candidate electron neutrino interaction. The scale shown applies to both images. The color is proportional to the charge collected.}
\end{figure}

The reconstructed vertex $dE/dx$ distribution for data after the final selection is shown in Figure \ref{fig:dedx}. One of the candidate electron neutrino interactions, among the 13 selected, is shown in Figure \ref{fig:nue}. The inset of Figure \ref{fig:dedx} shows the distribution of vertex $dE/dx$ for all reconstructed simulated electrons in the defined fiducial volume separated according to interaction mechanism [quasi-elastic, resonant, and deep inelastic scattering (DIS)]. Notably, the tail of the distribution is composed almost entirely of DIS interactions, an important contribution for the few-GeV neutrinos observed by ArgoNeuT.  

\begin{figure}[tbp]
\centering 
\includegraphics[width=.45\textwidth,trim={10 10 10 10}]{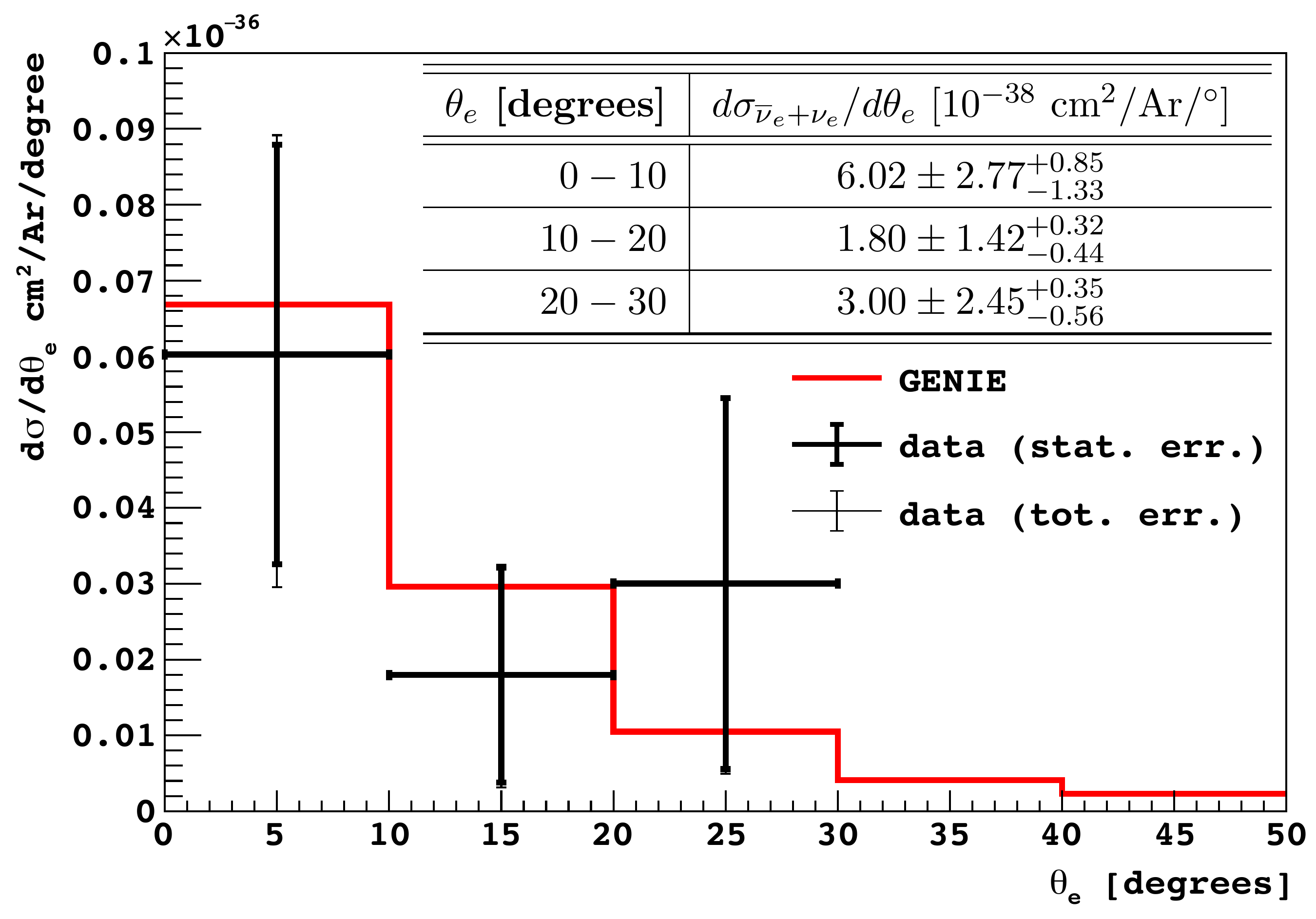}
\caption{\label{fig:angle} The ArgoNeuT $\nu_e + \overline{\nu}_e$ CC differential cross section for electron/positron angle with respect to the neutrino beam compared to the GENIE prediction.}
\end{figure}

ArgoNeuT cannot sign-select for neutrino type, and significant contributions of both $\nu_e$ and $\overline{\nu}_e$ interactions are expected in NuMI low-energy antineutrino mode data. Thus, we define a flux-averaged total cross section such that it is a combination of  $\nu_e$ and $\overline{\nu}_e$~\cite{uboone_nue}: $\sigma_{\nu_e + \overline{\nu}_e} = (N - B)/(\epsilon N_\mathrm{Ar} (\Phi_{\nu_e} + \Phi_{\overline{\nu}_e}))$ where $N$ is the number of events selected in data, $B$ is the number of background events in simulation, $\epsilon$ is the selection efficiency, $N_\mathrm{Ar}$ is the number of argon targets, and $\Phi$ is integrated flux. The $\nu_e$ and $\overline{\nu}_e$ fluxes can be found in the supplemental material of Ref. \cite{numi_flux2}. Using this convention, we extract a total cross section of $(1.04 \pm 0.38~\mathrm{(stat.)} ^{+0.15}_{-0.23}~\mathrm{(syst.)}) \times 10^{-36} ~\mathrm{cm}^2$ on argon ($\langle E_{\nu_e} \rangle = 10.5$ GeV and $\langle E_{\bar{\nu}_e} \rangle = 4.3$ GeV) consistent with the GENIE expectation ($11.2^{+0.4}_{-1.4}$ signal events and $3.0^{+2.0}_{-1.3}$ total background events). Further, we report a differential cross section in Figure \ref{fig:angle}, again as a combination of $\nu_e$ and $\overline{\nu}_e$ defined in this way: $d\sigma(\theta_{e,i})/d\theta_e = (N_i - B_i)/(\epsilon_i \Delta\theta_{e,i} N_\mathrm{Ar} (\Phi_{\nu_e} + \Phi_{\overline{\nu}_e}))$. Uncertainties associated with the GENIE modeling contribute most to the systematic uncertainties. Importantly, the interpretation of these results, for example in comparisons to model predictions and event generators, \textit{requires} the consideration of both the detailed $\nu_e$ and $\overline{\nu}_e$ fluxes simultaneously~\cite{numi_flux2}.

There are several notable factors that impact the performance achieved in this analysis ($10.5^{+0.6}_{-0.5}$\% efficiency with $78.9^{+8.1}_{-11.8}$\% purity). While the efficiency is sufficient for exploring the  data-driven classification techniques and performing the measurements reported here, it is limited by ArgoNeuT's intrinsic reconstruction capabilities. First, ArgoNeuT's size is such that EM shower containment is a rarity, which leads to difficulty in event classification. Poor track containment, in general, also affects vertex reconstruction and event classification. Second, the signal selection in ArgoNeuT is necessarily very strict since we cannot move sufficiently far away from the detector walls in the fiducial volume definition to reduce background events produced by interactions external to the active volume of the detector, most notably single gammas. With improvements to these issues, as expected in future detectors, like DUNE, we expect a significant increase in inclusive $\nu_e/\overline{\nu}_e$ CC signal selection efficiency.


We have reported a total $\nu_e + \overline{\nu}_e $ cross section and a differential cross section in terms of electron/positron angle with respect to the incoming neutrino using the fully-automated reconstruction and analysis framework described above. These are the first measurements of electron neutrino scattering cross sections on argon. The results are statistics limited, further affected by the reconstruction efficiency in ArgoNeuT and the strict selection required to mitigate external backgrounds in the small detector. Furthermore, this is the first measurement of electron neutrino and antineutrino scattering using the same target nucleus and over the same energy range that will be used by the DUNE experiment.

The unique selection techniques outlined in this Letter are particularly useful for identifying $\nu_e/\overline{\nu}_e$ CC interactions among typical GeV-scale neutrino backgrounds, including events involving gamma-induced showers and/or containing multiple tracks and complicated topologies. Our approach considers the topology and charge distribution of the entire candidate electron shower, rather than relying solely on the traditional near-vertex EM shower conversion gap and $dE/dx$ information, which can be obscured/ambiguous at the GeV-scale. Further development of calorimetry-based techniques for signal classification is critical to inform and direct machine learning-based image classification methods currently at the forefront of pattern recognition technology~\cite{deep_learning_microboone, deep_learning_nature, wirecell}.

This manuscript has been authored by Fermi Research Alliance, LLC under Contract No. DE-AC02-07CH11359 with the U.S. Department of Energy, Office of Science, Office of High Energy Physics. We gratefully acknowledge the cooperation of the MINOS Collaboration in providing their data for use in this analysis. We wish to acknowledge the support of Fermilab, the Department of Energy, and the National Science Foundation in ArgoNeuT's construction, operation, and data analysis.

\clearpage

\onecolumngrid

\section{Supplementary Material}

\noindent The following pages show the full set of 13 candidate events selected in the ArgoNeuT automated search for $\nu_e/\overline{\nu}_e$ charged current interactions. The induction plane (left) and collection plane (right) views are both shown. These images scale to approximately 90~cm in the wire direction, which increases along the beam direction, and approximately 62 cm in the drift direction (measured in time sample number, noting that the time range is larger than the size of the detector in the drift direction, 47~cm). The color is proportional to the charge collected. Coherent noise is present in some images around approximately wire 200-250, sample number 1750. Other images contain bursts of charge due to activity on the opposite side of the wire planes in approximately wire 230, sample number 0-500 region.

The last three event displays show the obvious-by-eye background interactions spuriously identified as signal in the final selection. The first background event shows a single gamma-induced shower separated from the interaction vertex. The second is a through-going muon, and the third is track-like. These backgrounds and the event rate are consistent with expectations from simulation. 

\clearpage

\begin{figure}[h] 
\includegraphics[width=0.49\textwidth]{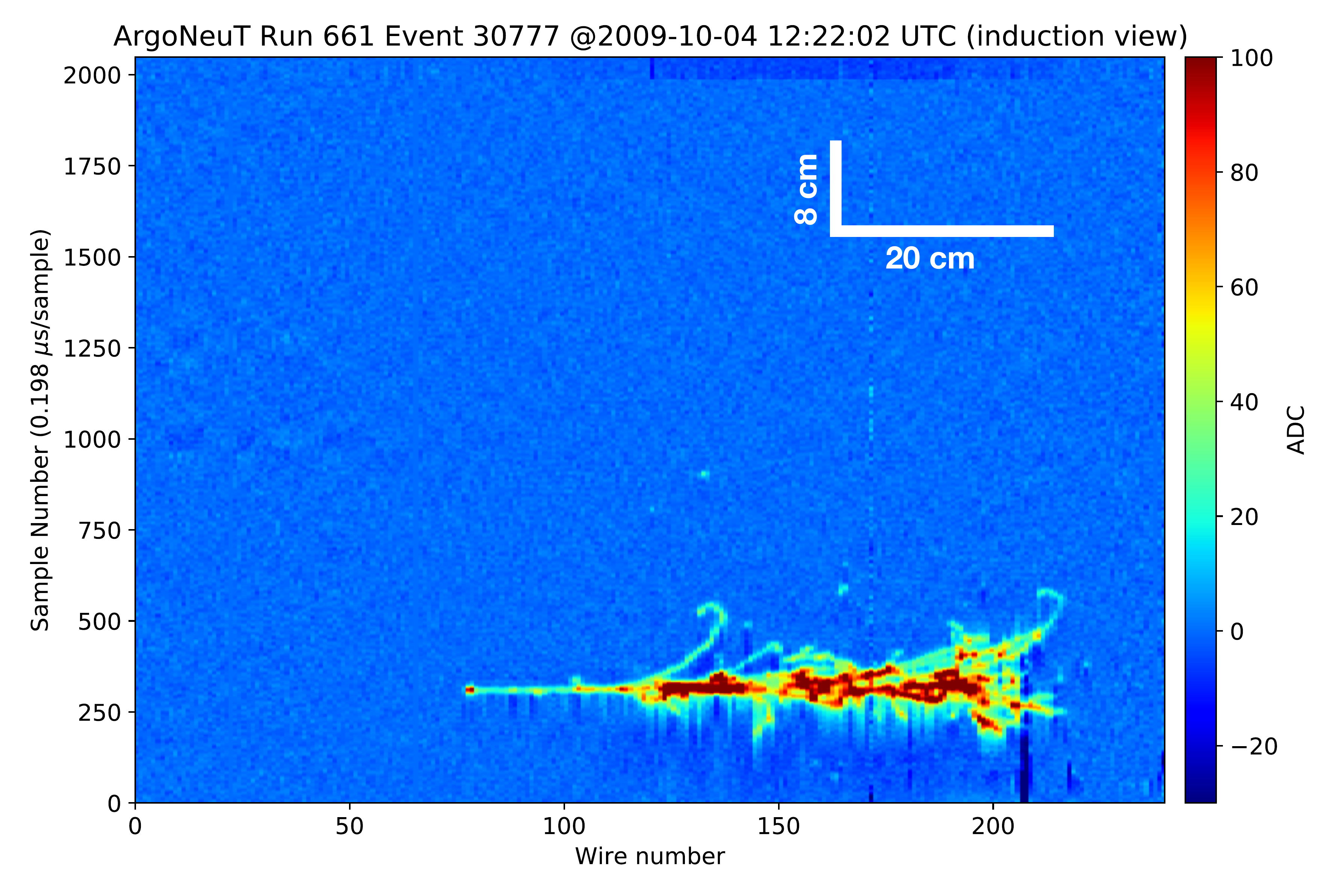}
\includegraphics[width=0.49\textwidth]{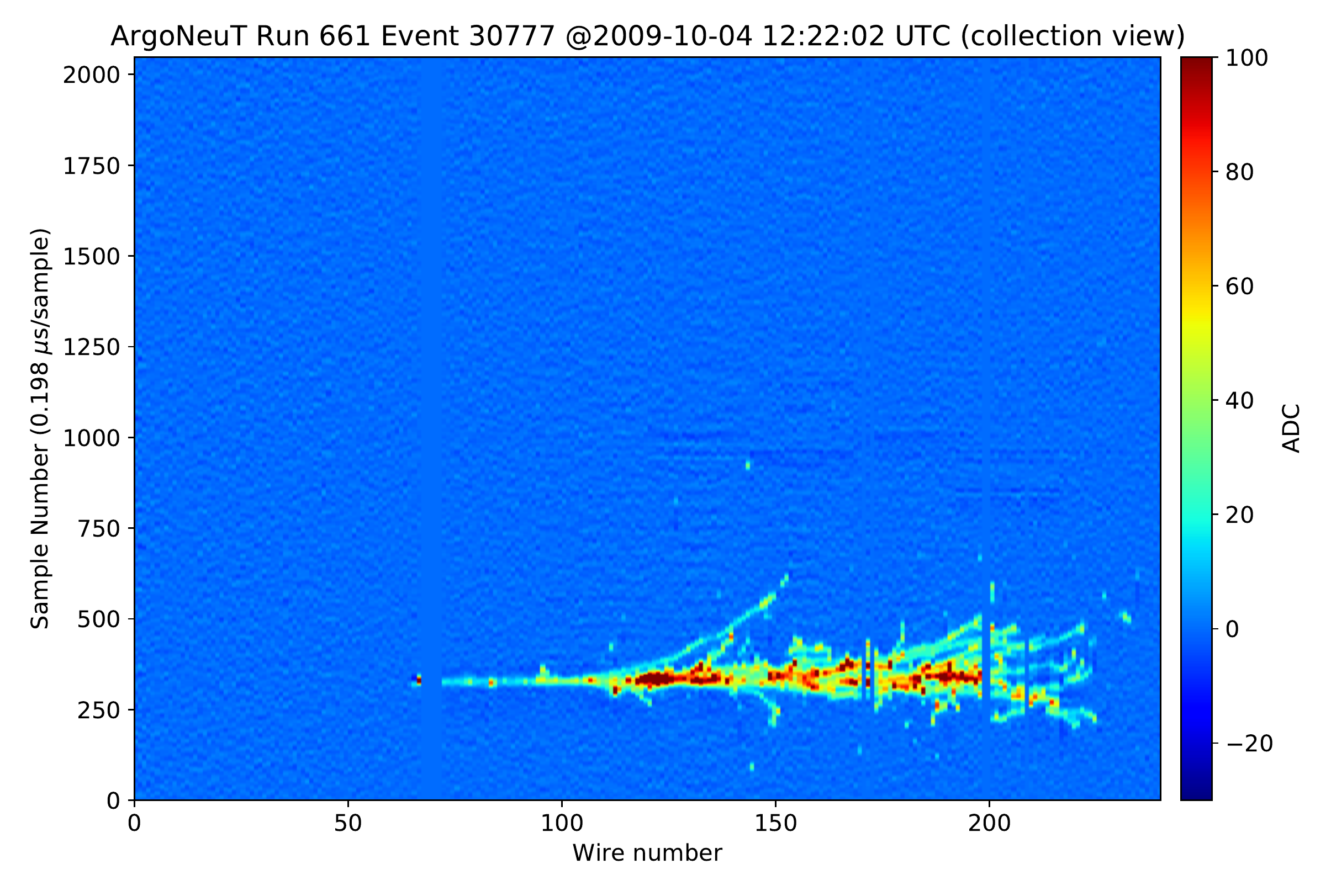}
\end{figure}    

\begin{figure*}[th!]
\centering 
\includegraphics[width=.49\textwidth]{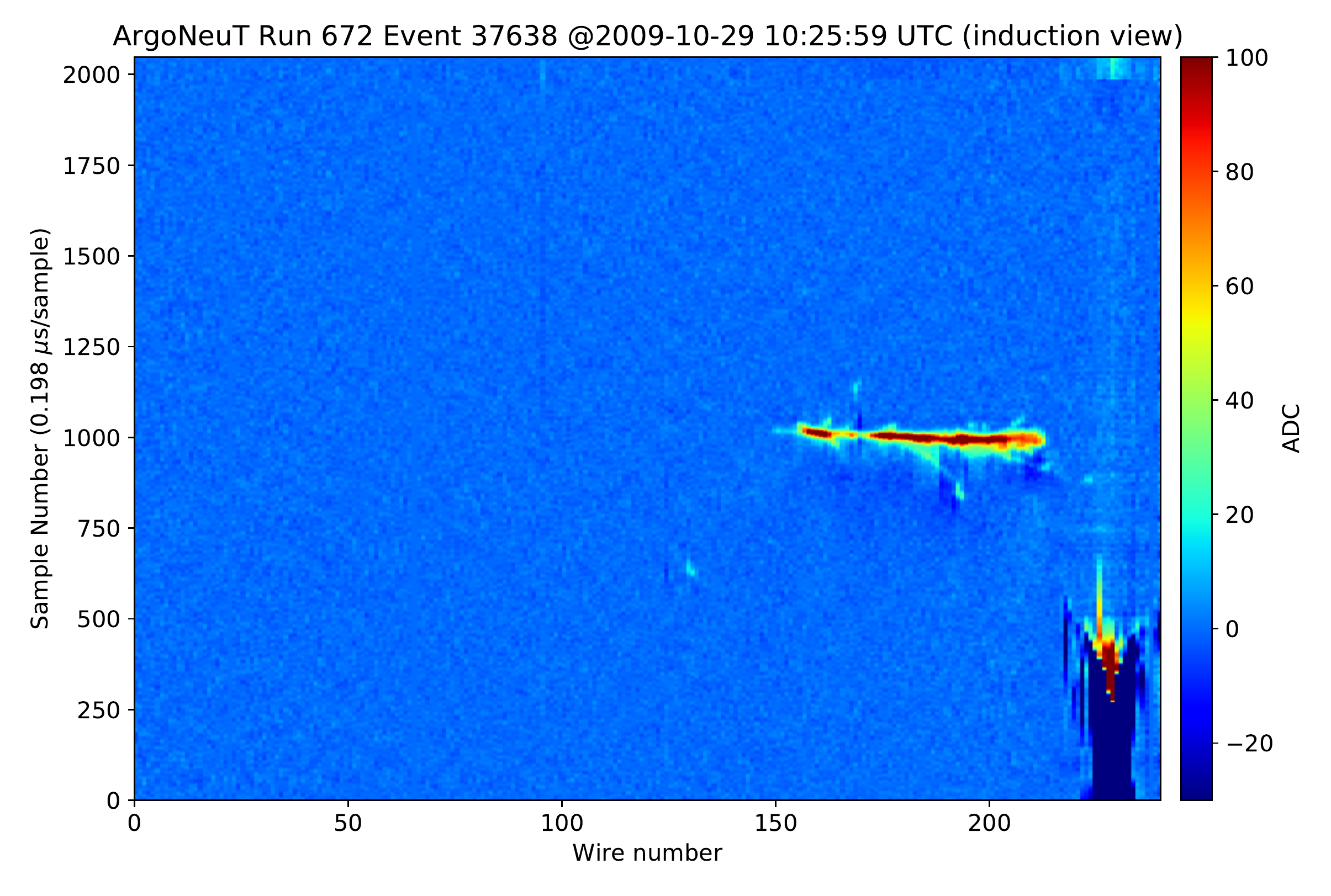} 
\includegraphics[width=.49\textwidth]{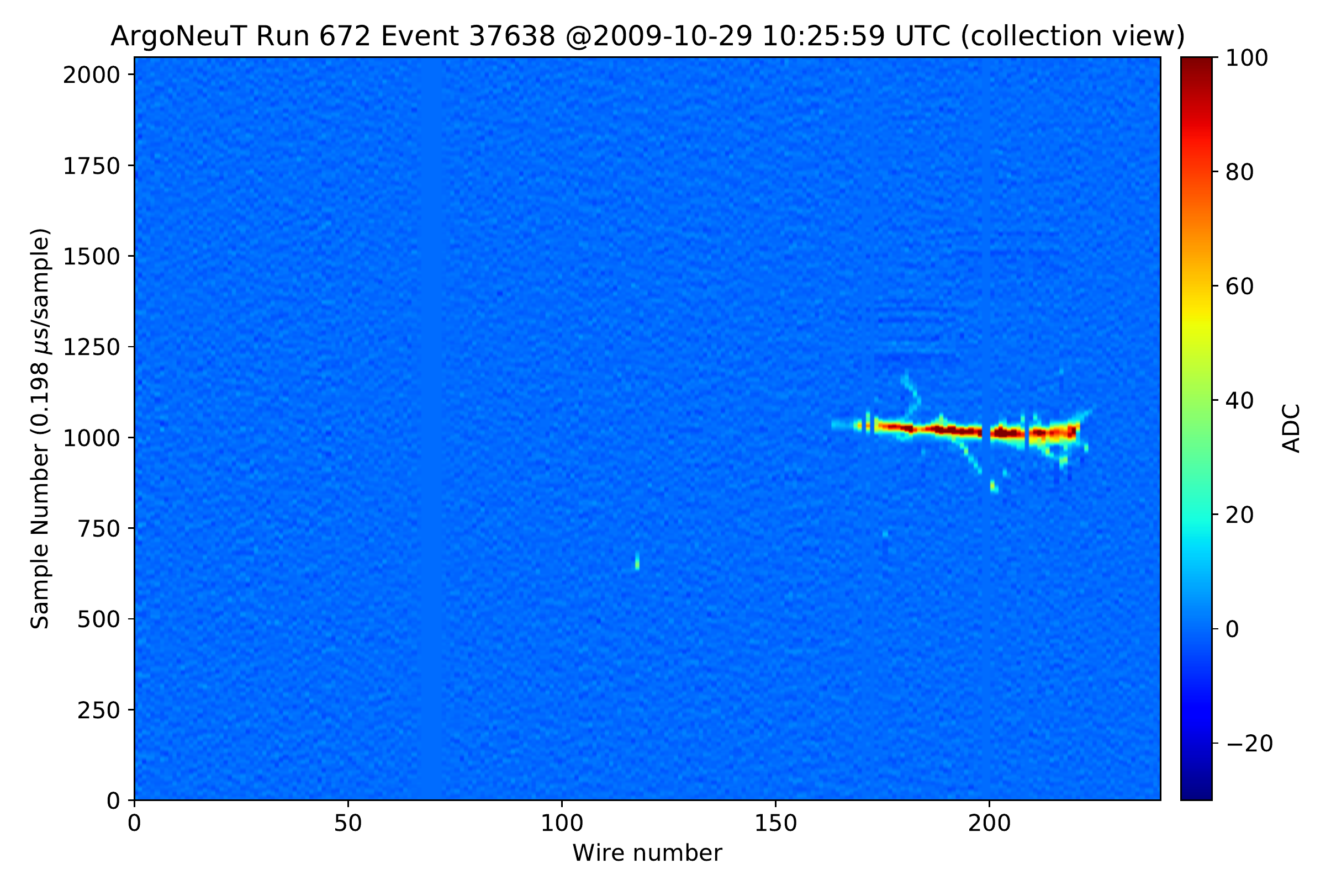} 
\end{figure*}

\begin{figure*}[th!]
\centering 
\includegraphics[width=.49\textwidth]{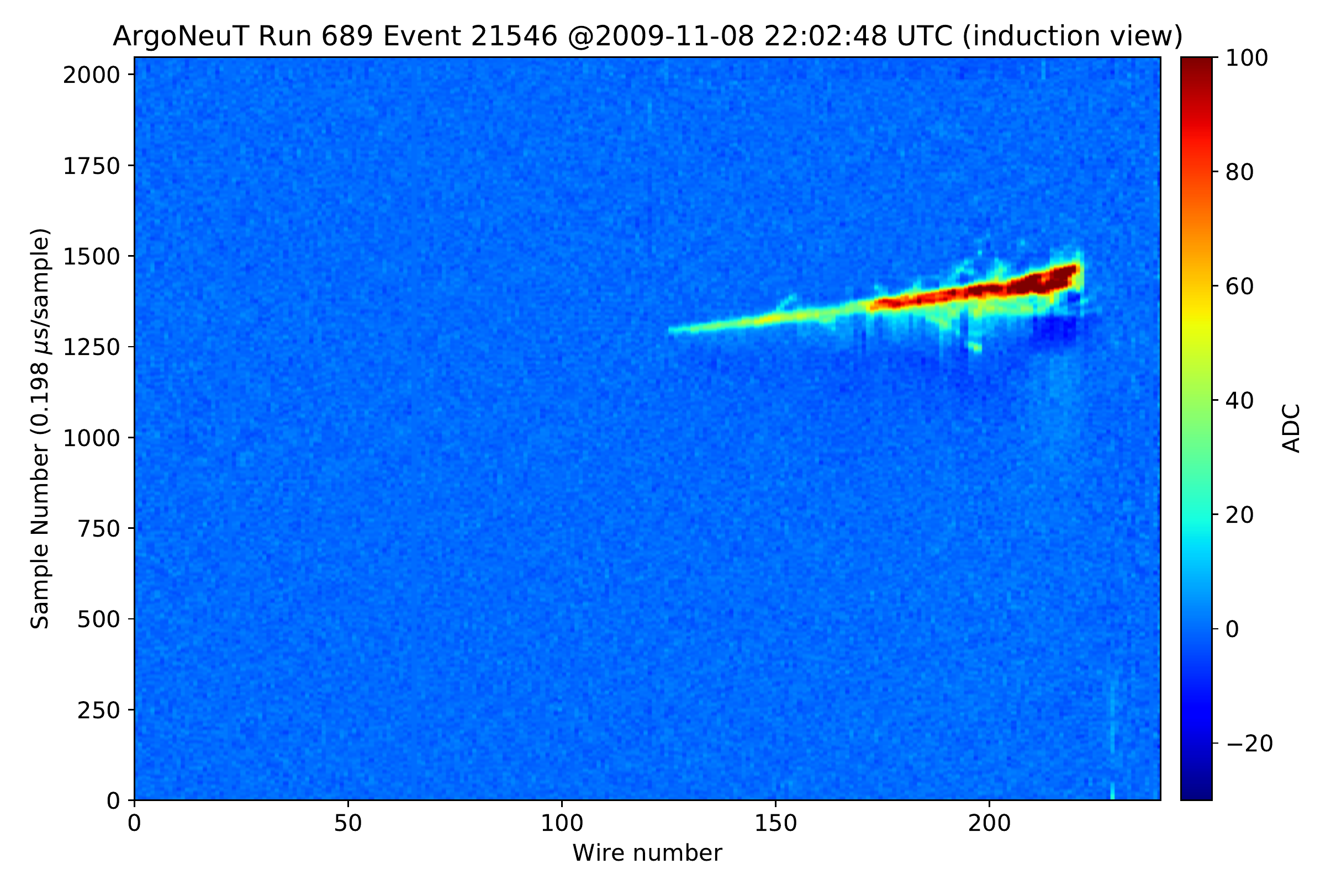} 
\includegraphics[width=.49\textwidth]{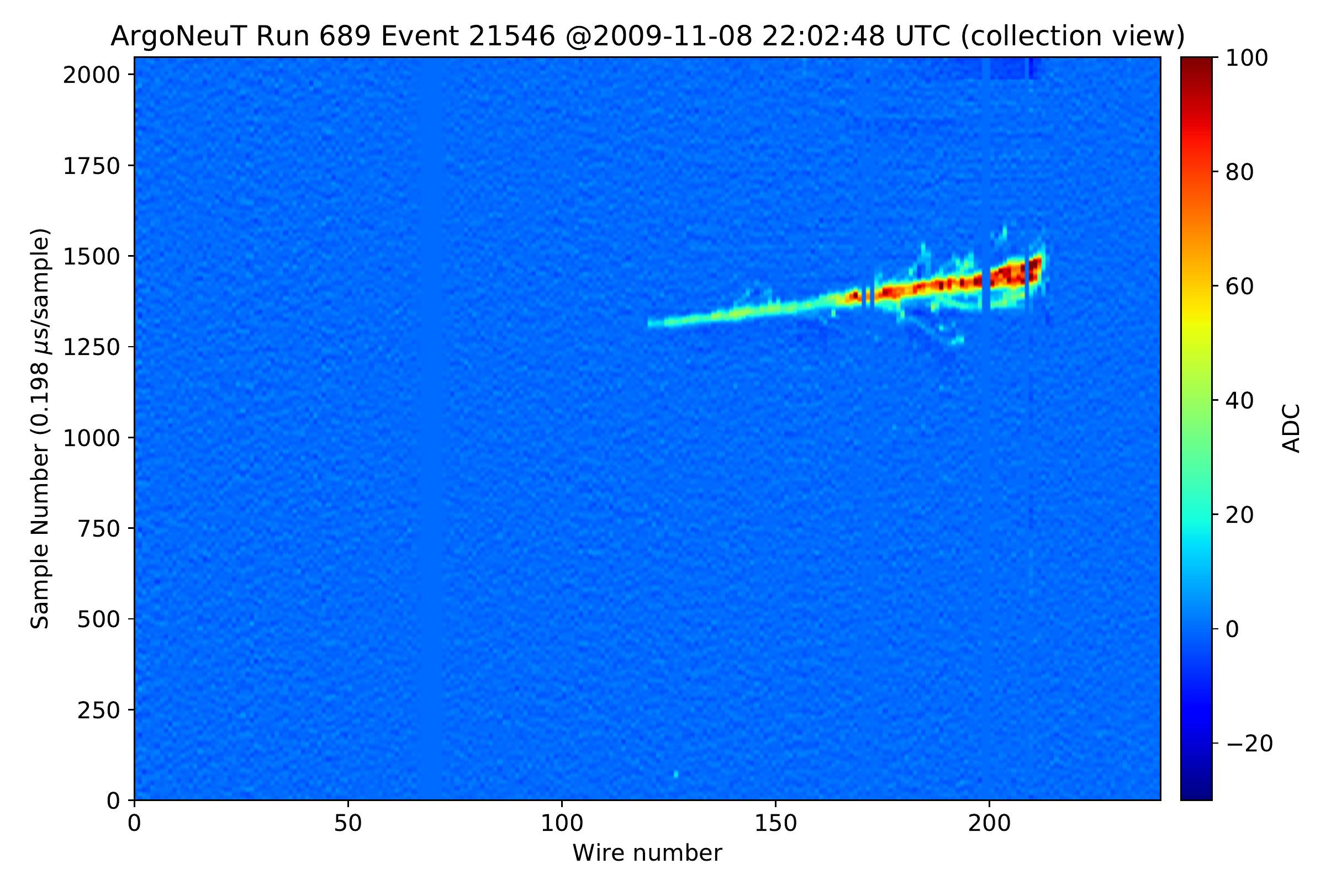} 
\end{figure*}

\begin{figure*}[th!]
\centering 
\includegraphics[width=.49\textwidth]{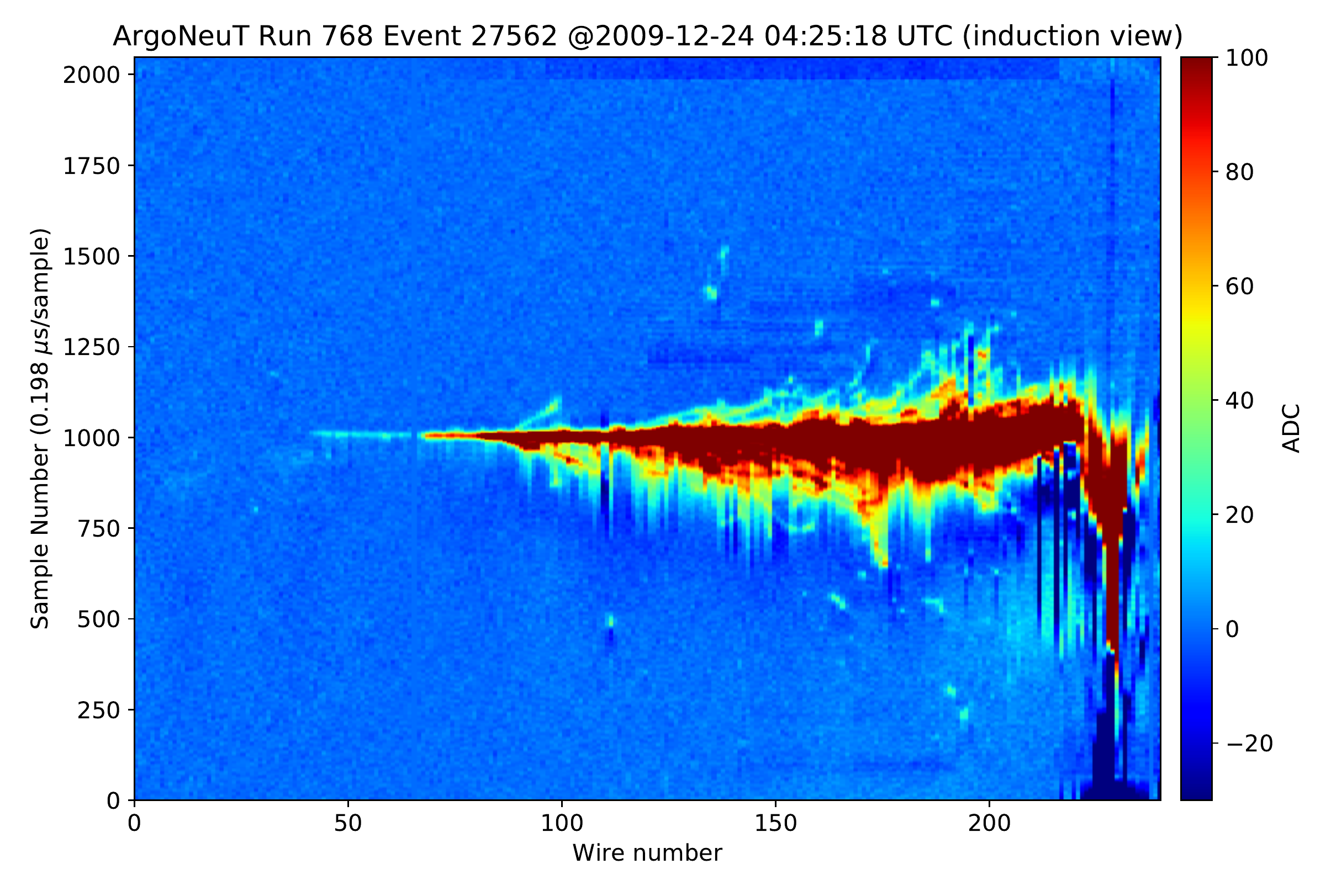} 
\includegraphics[width=.49\textwidth]{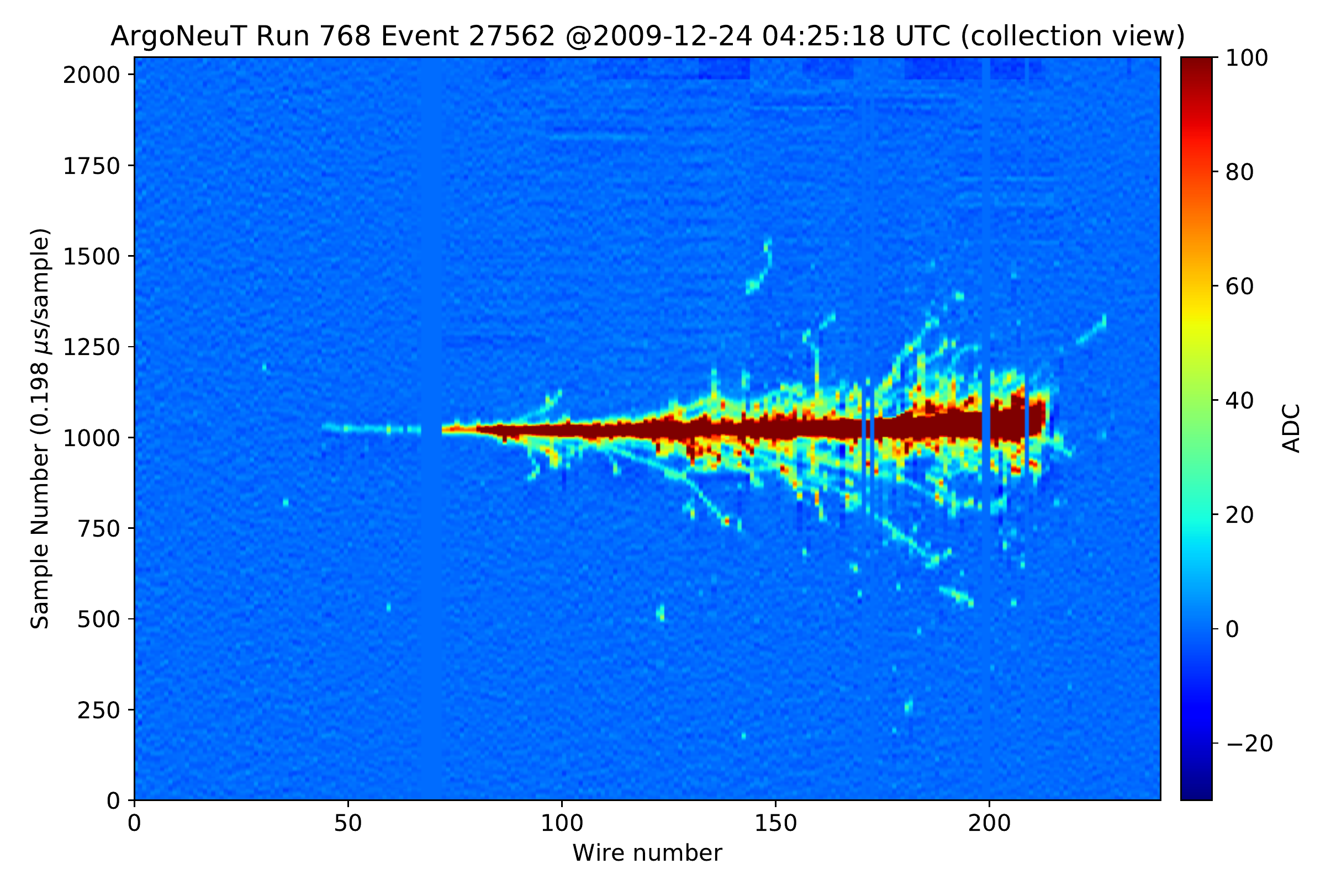} 
\end{figure*}

\begin{figure*}[th!]
\centering 
\includegraphics[width=.49\textwidth]{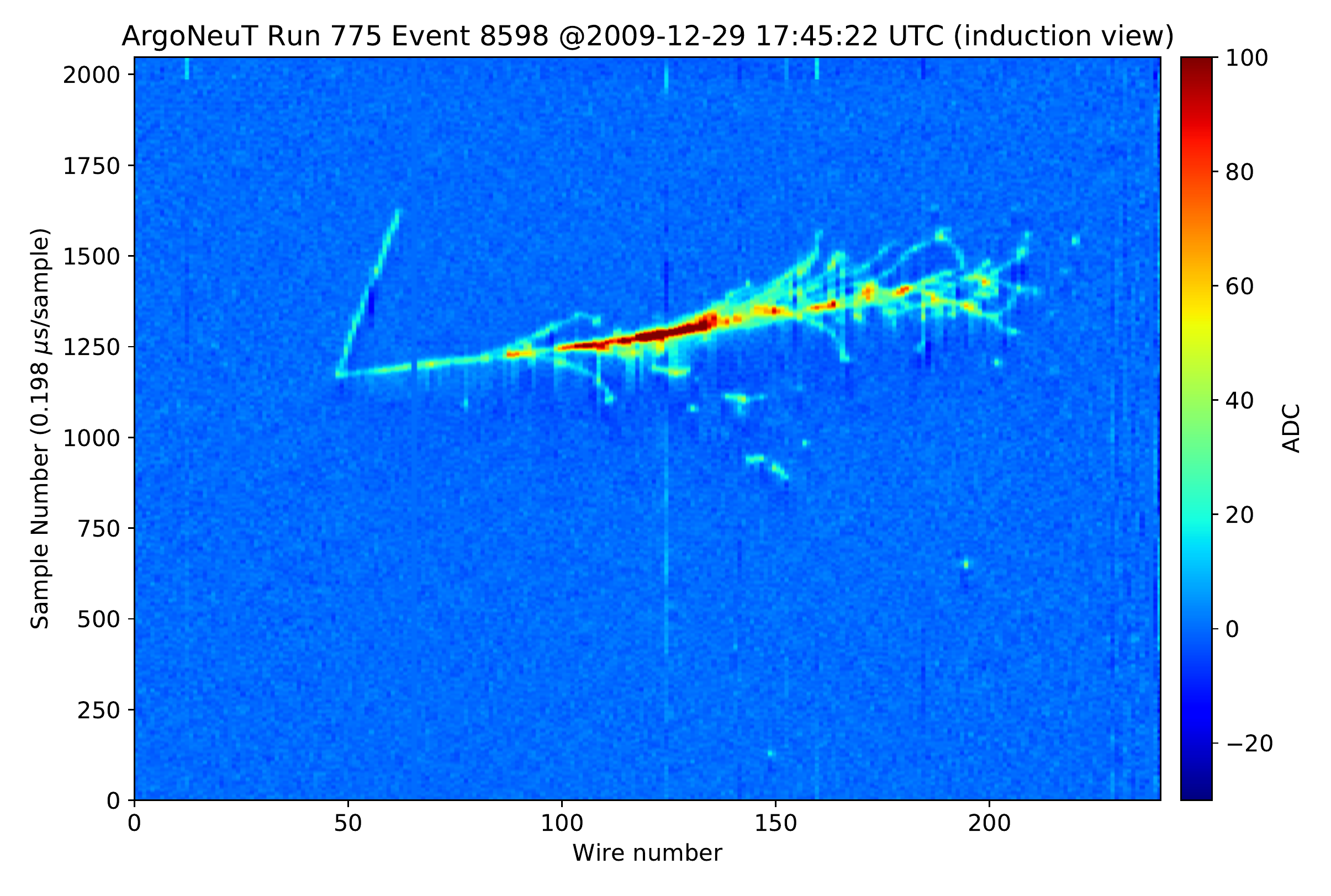} 
\includegraphics[width=.49\textwidth]{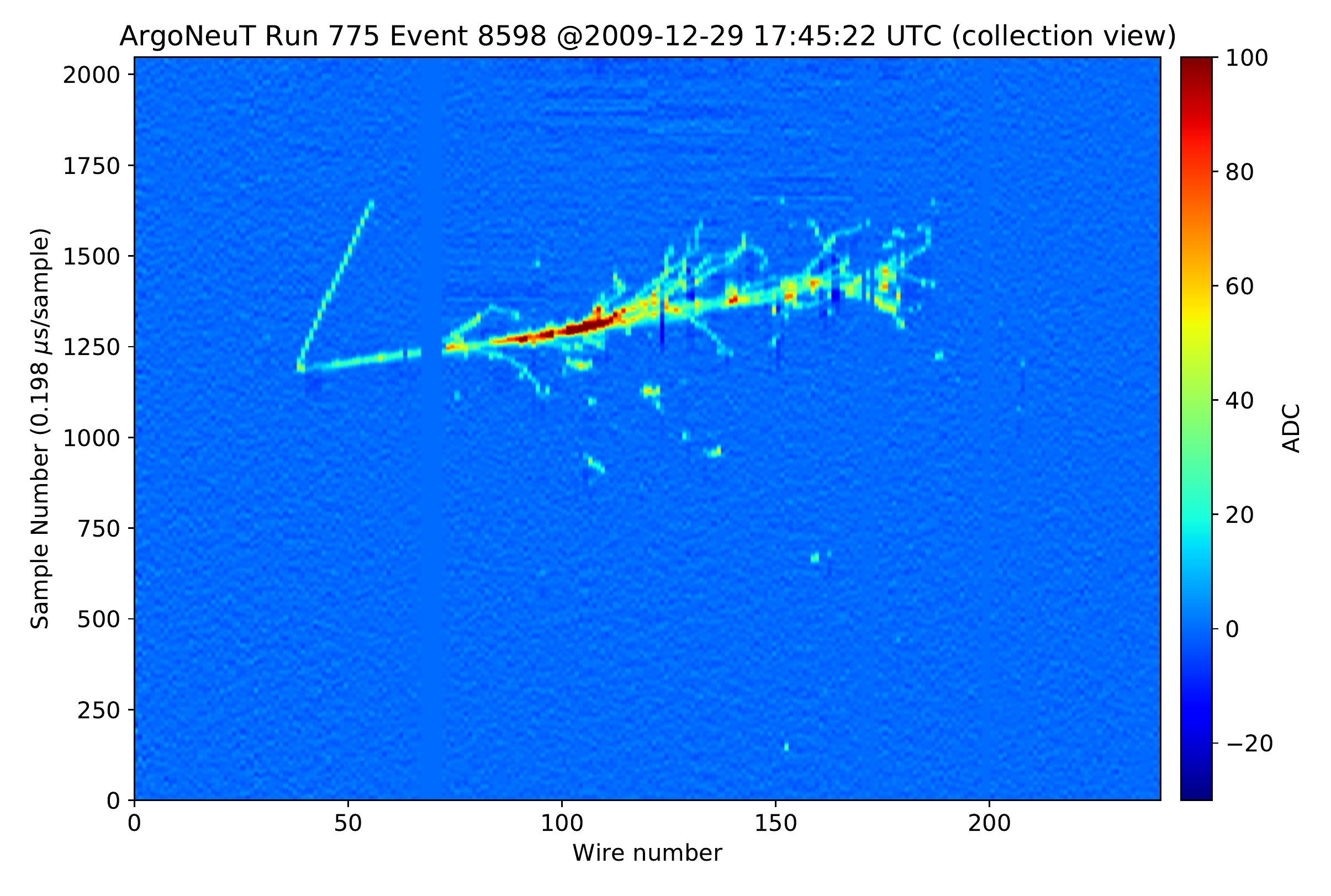} 
\end{figure*}

\begin{figure*}[th!]
\centering 
\includegraphics[width=.49\textwidth]{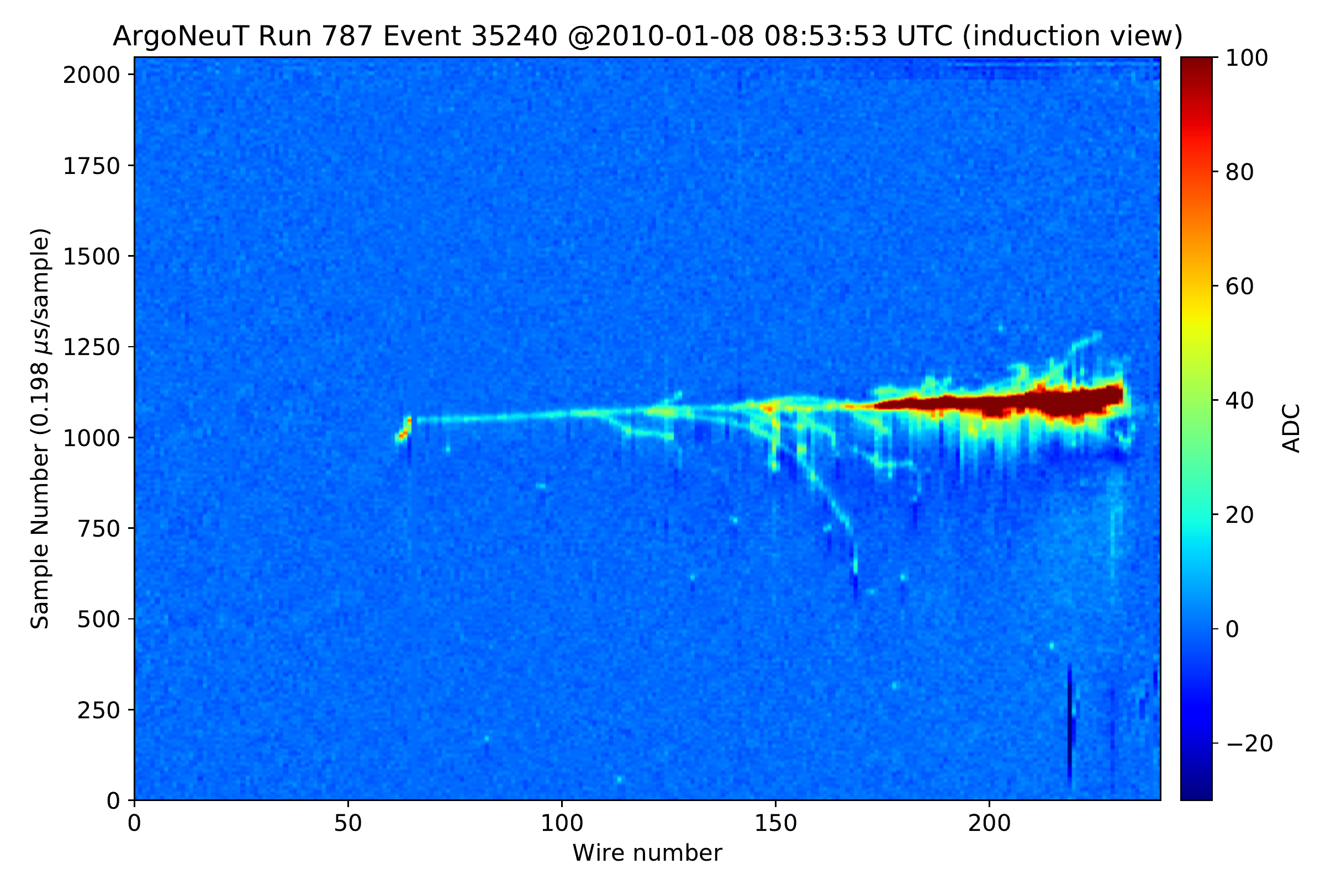} 
\includegraphics[width=.49\textwidth]{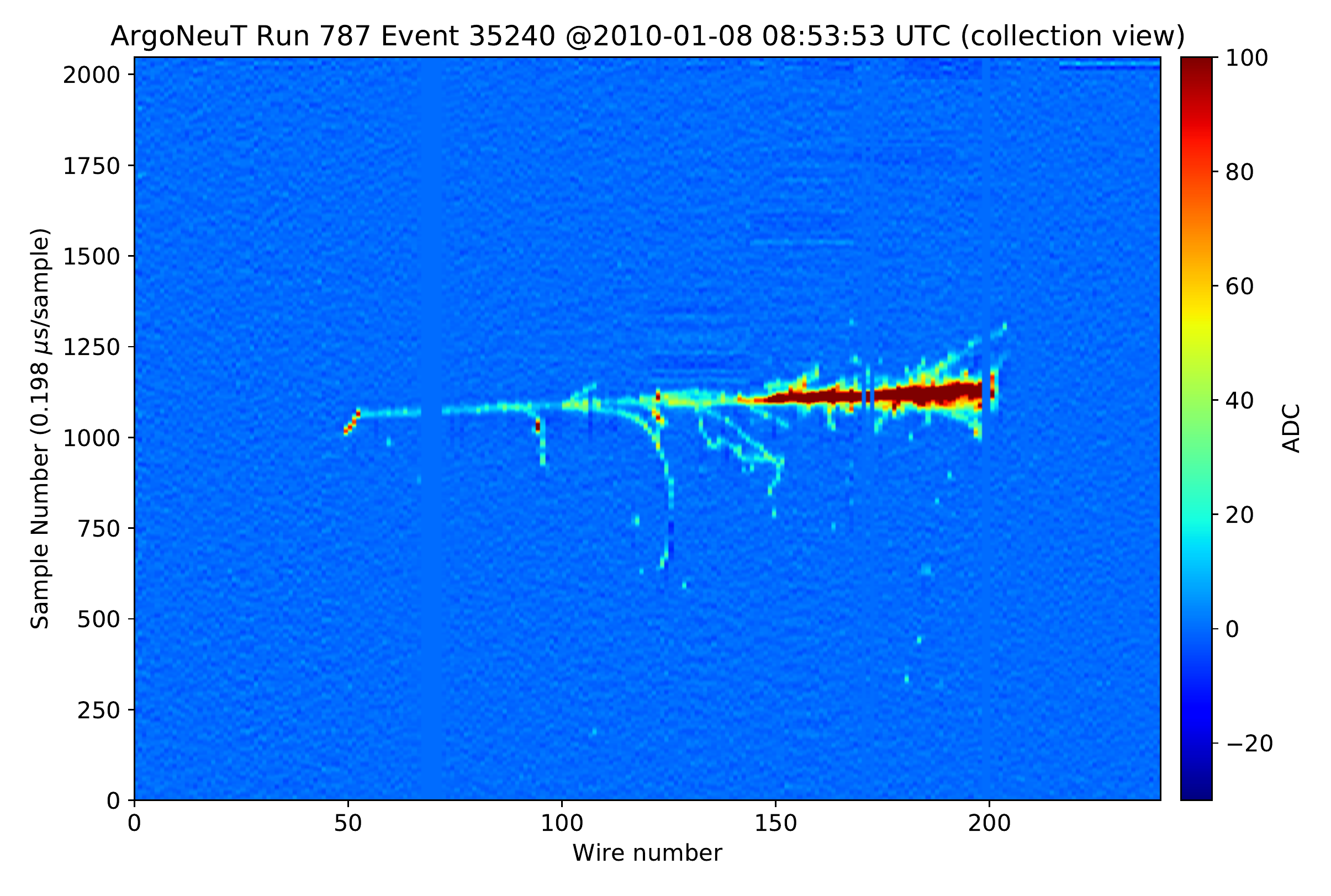} 
\end{figure*}

\begin{figure*}[th!]
\centering 
\includegraphics[width=.49\textwidth]{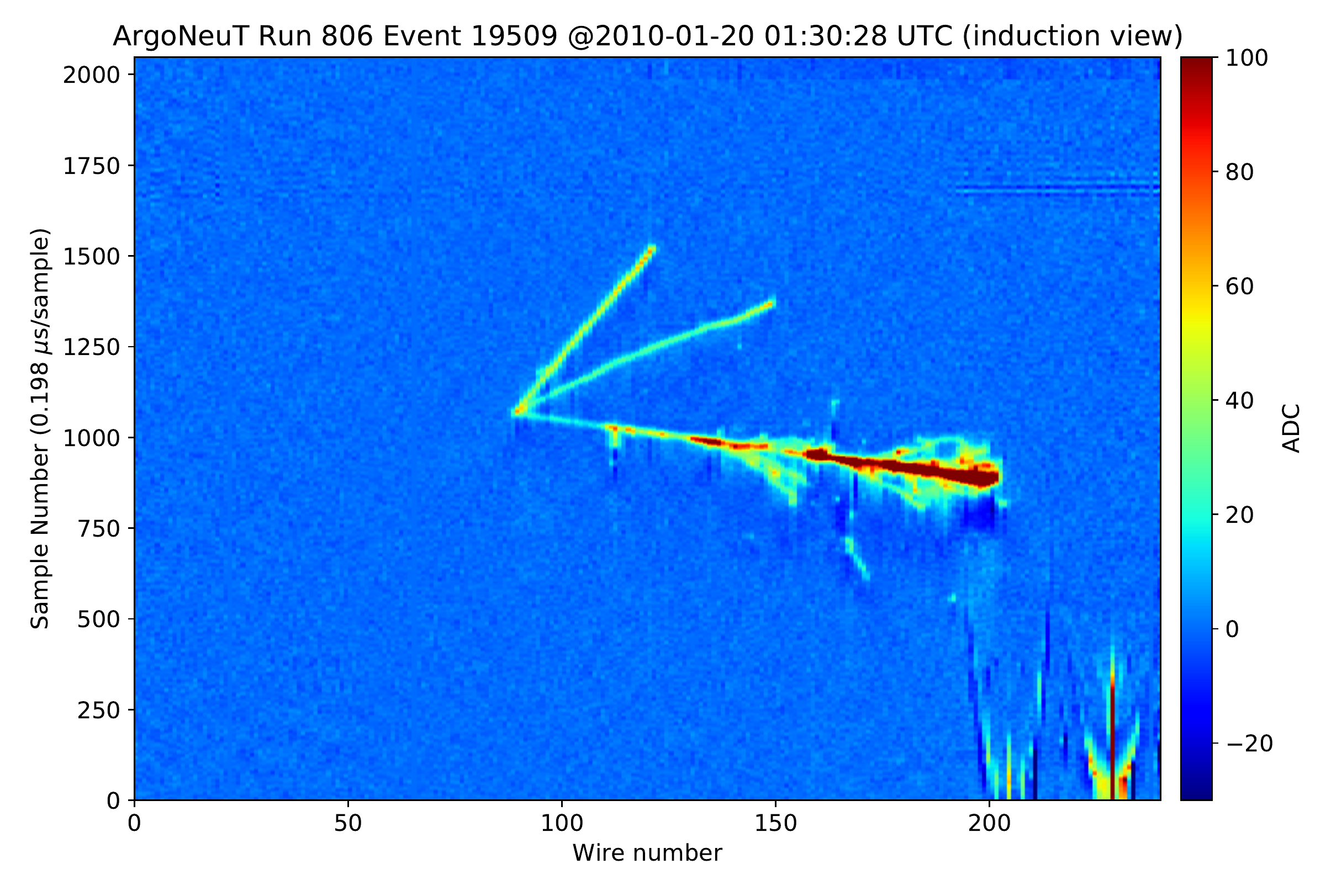} 
\includegraphics[width=.49\textwidth]{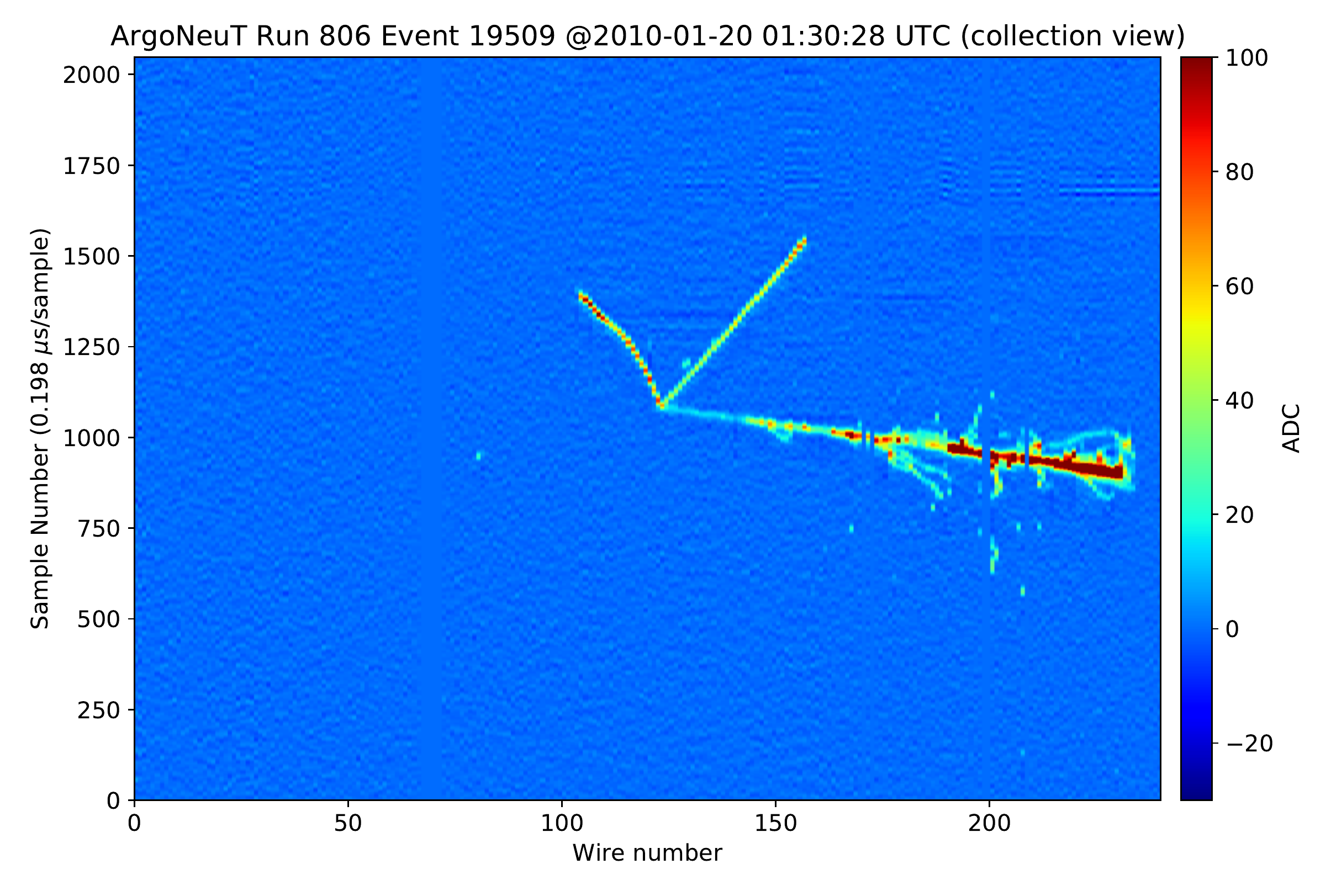} 
\end{figure*}

\begin{figure*}[th!]
\centering 
\includegraphics[width=.49\textwidth]{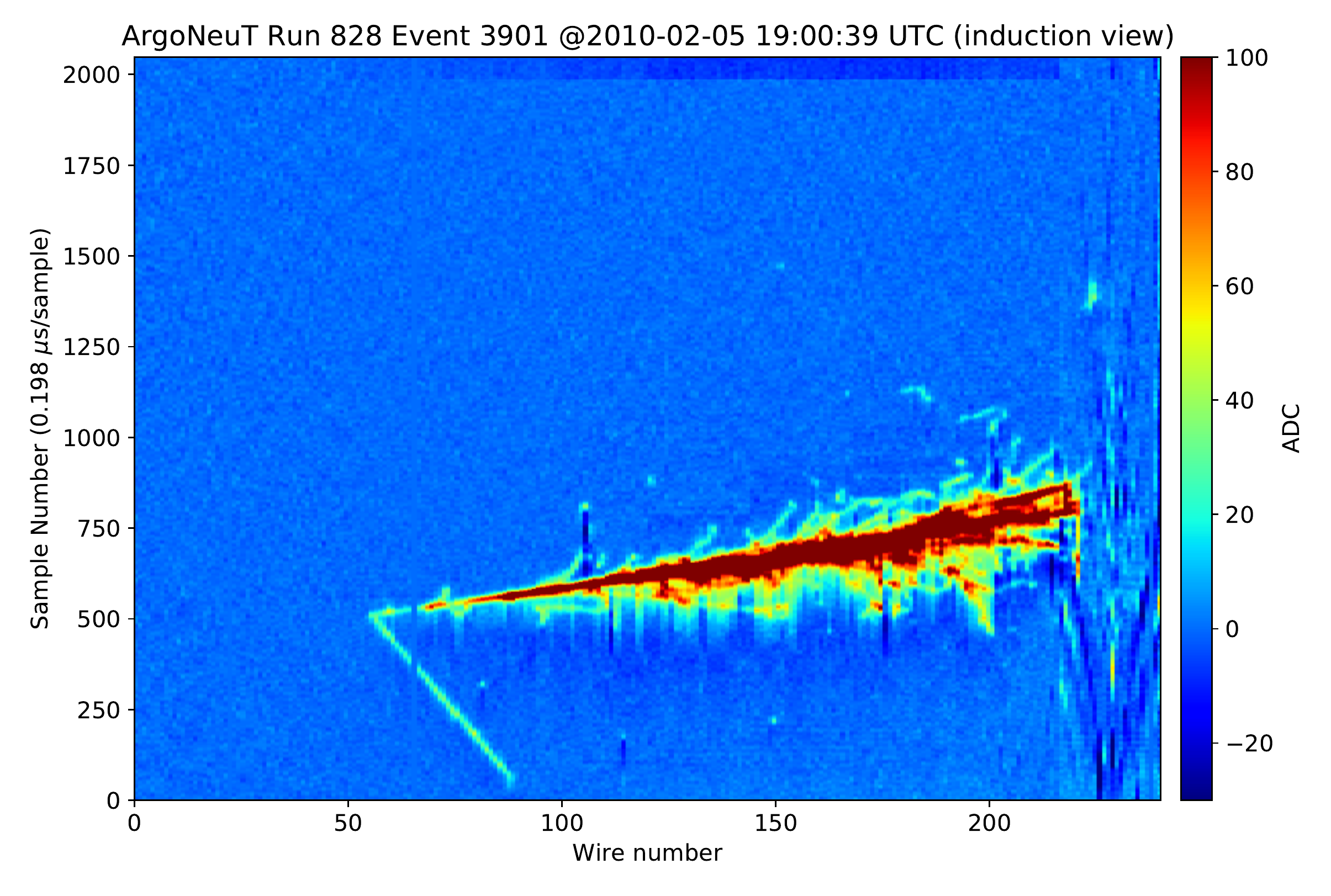} 
\includegraphics[width=.49\textwidth]{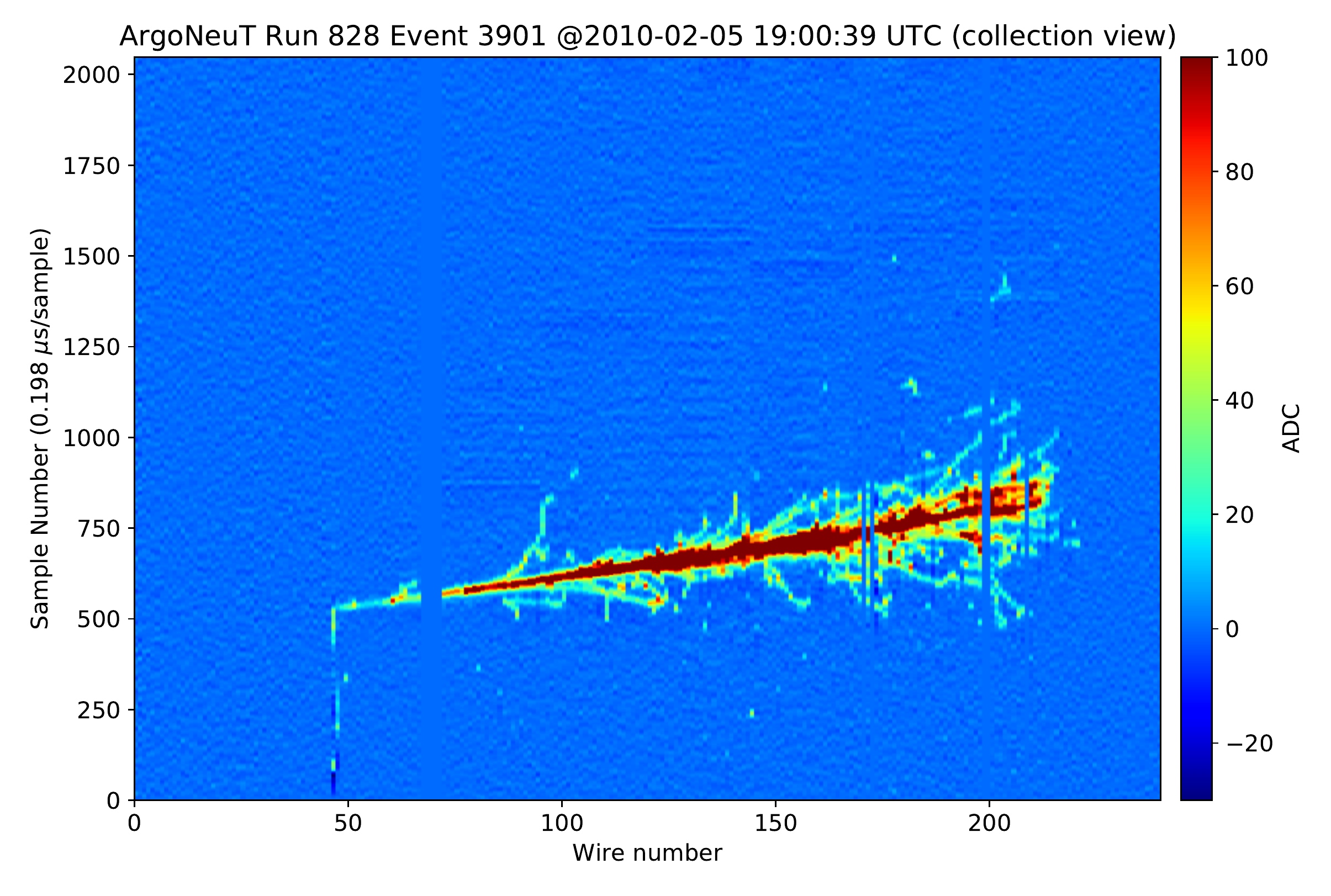} 
\end{figure*}

\begin{figure*}[th!]
\centering 
\includegraphics[width=.49\textwidth]{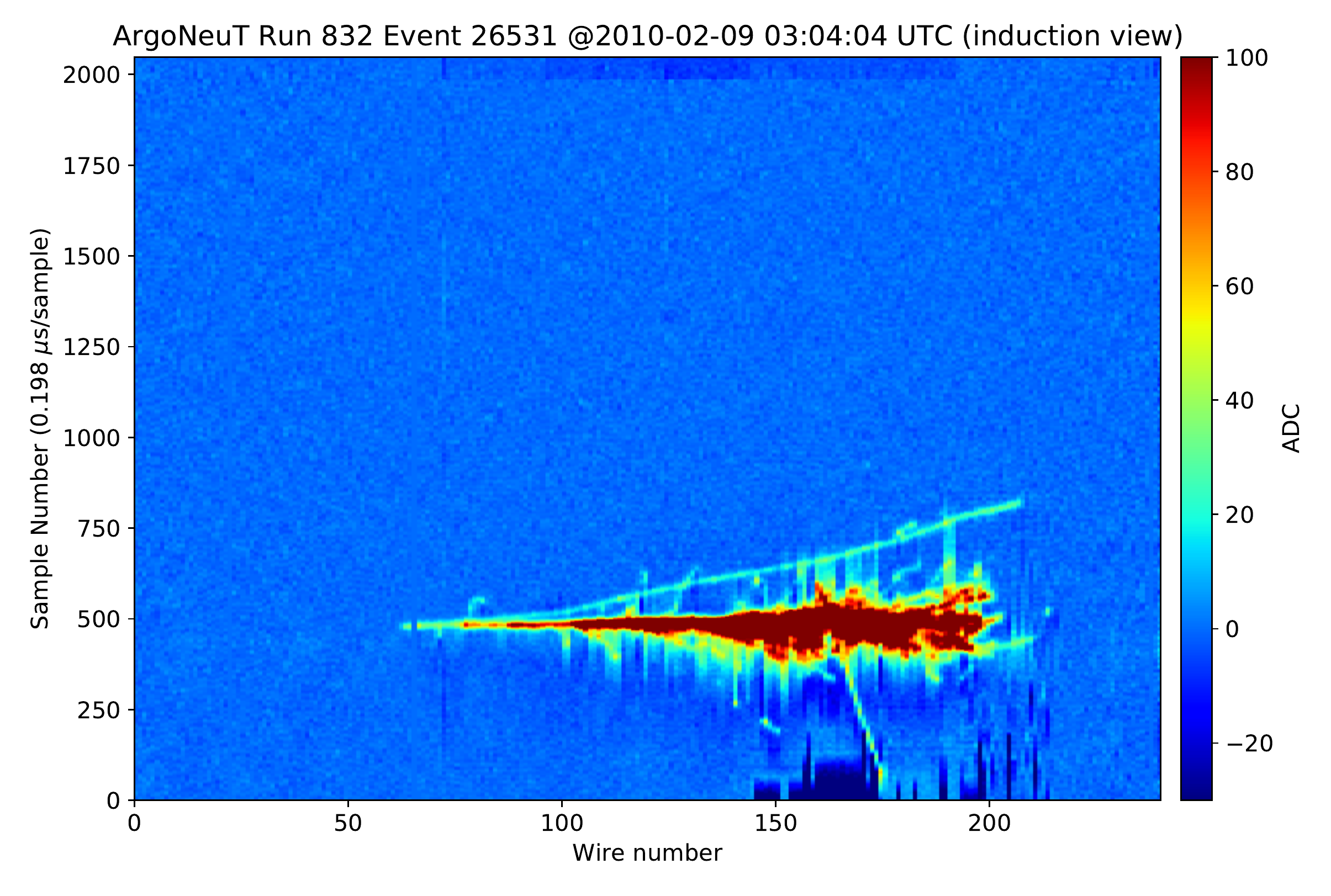} 
\includegraphics[width=.49\textwidth]{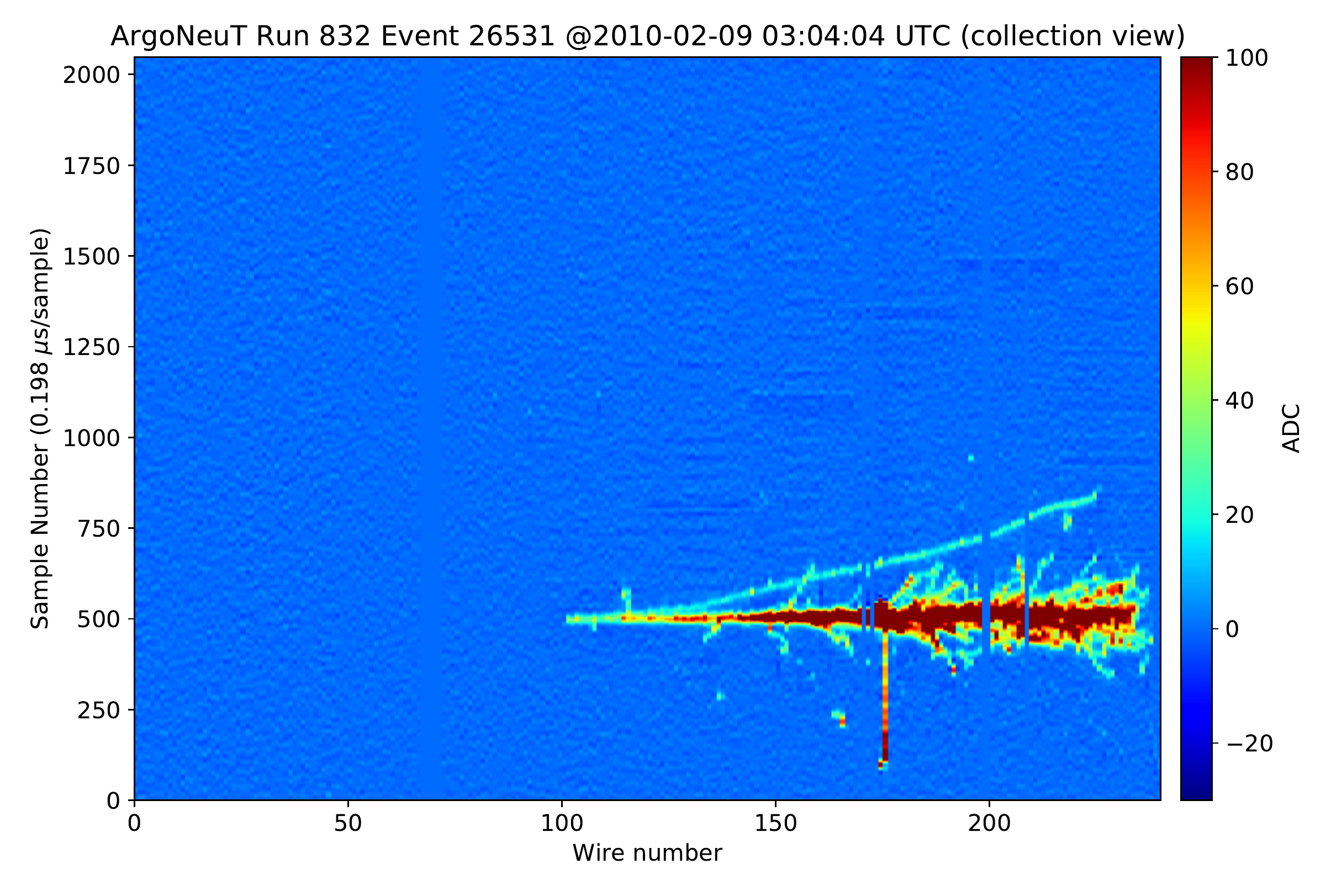} 
\end{figure*}

\begin{figure*}[th!]
\centering 
\includegraphics[width=.49\textwidth]{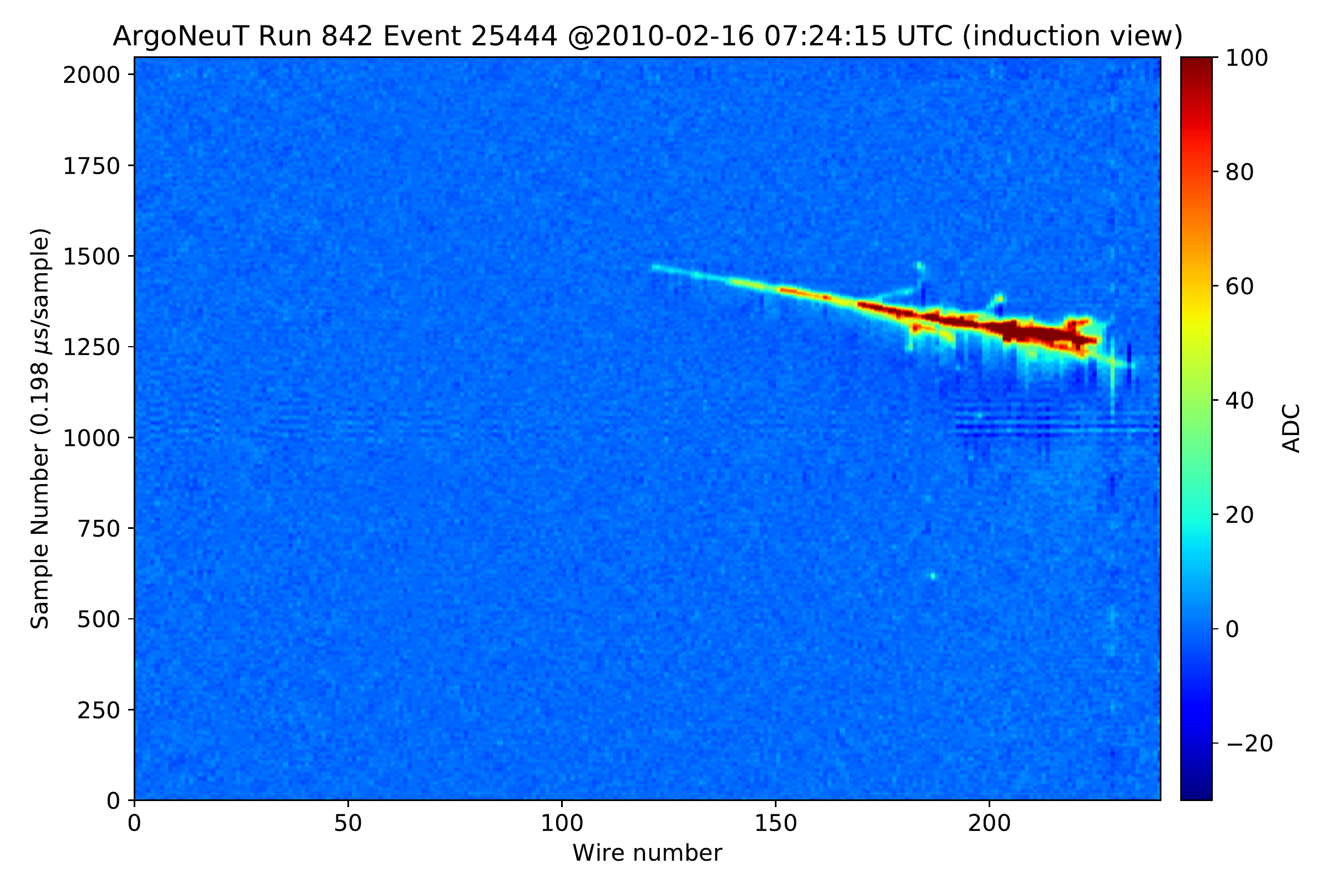} 
\includegraphics[width=.49\textwidth]{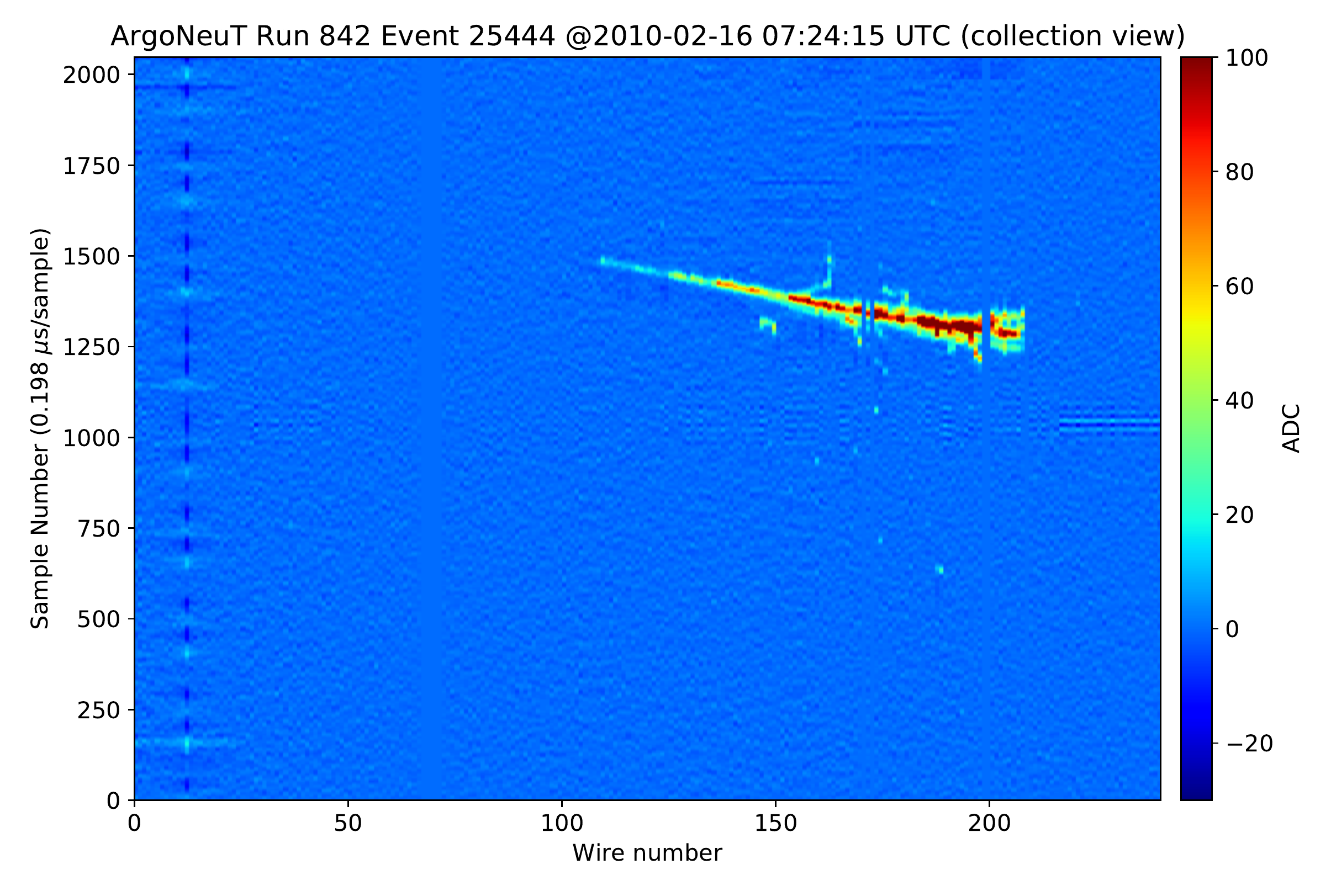} 
\end{figure*}
\clearpage
\begin{figure*}[th!]
\centering 
\includegraphics[width=.49\textwidth]{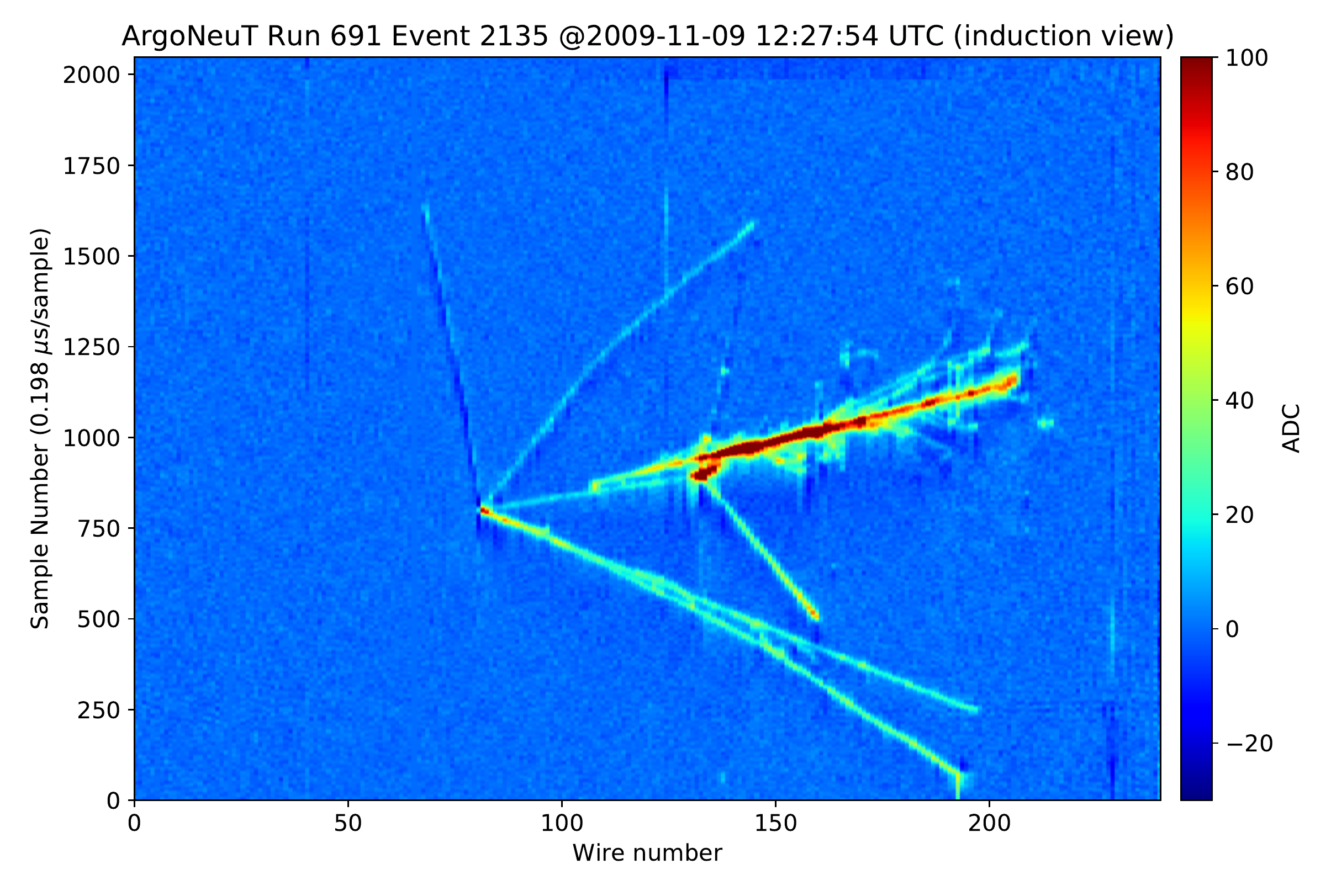} 
\includegraphics[width=.49\textwidth]{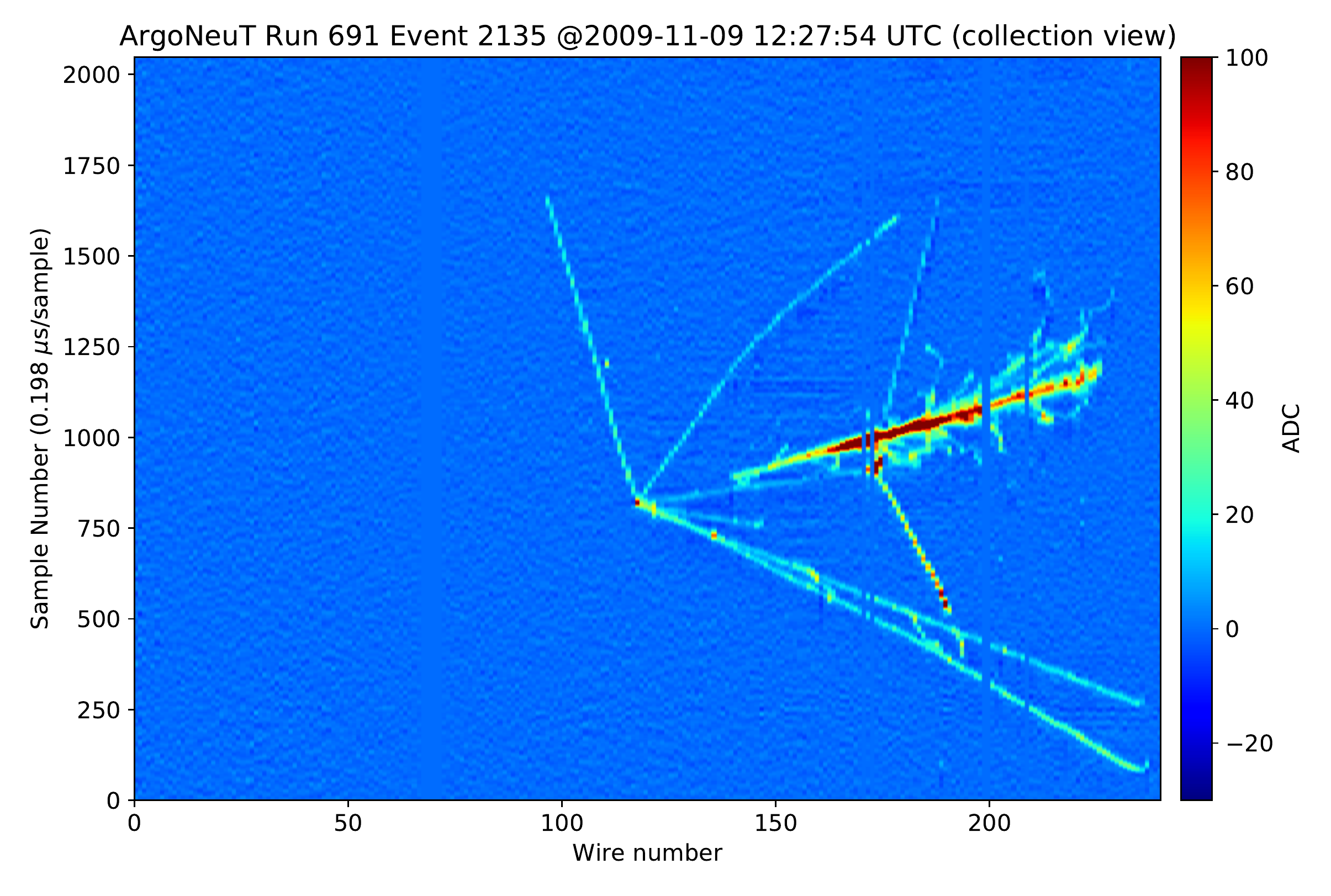} 
\end{figure*}

\begin{figure*}[th!]
\centering 
\includegraphics[width=.49\textwidth]{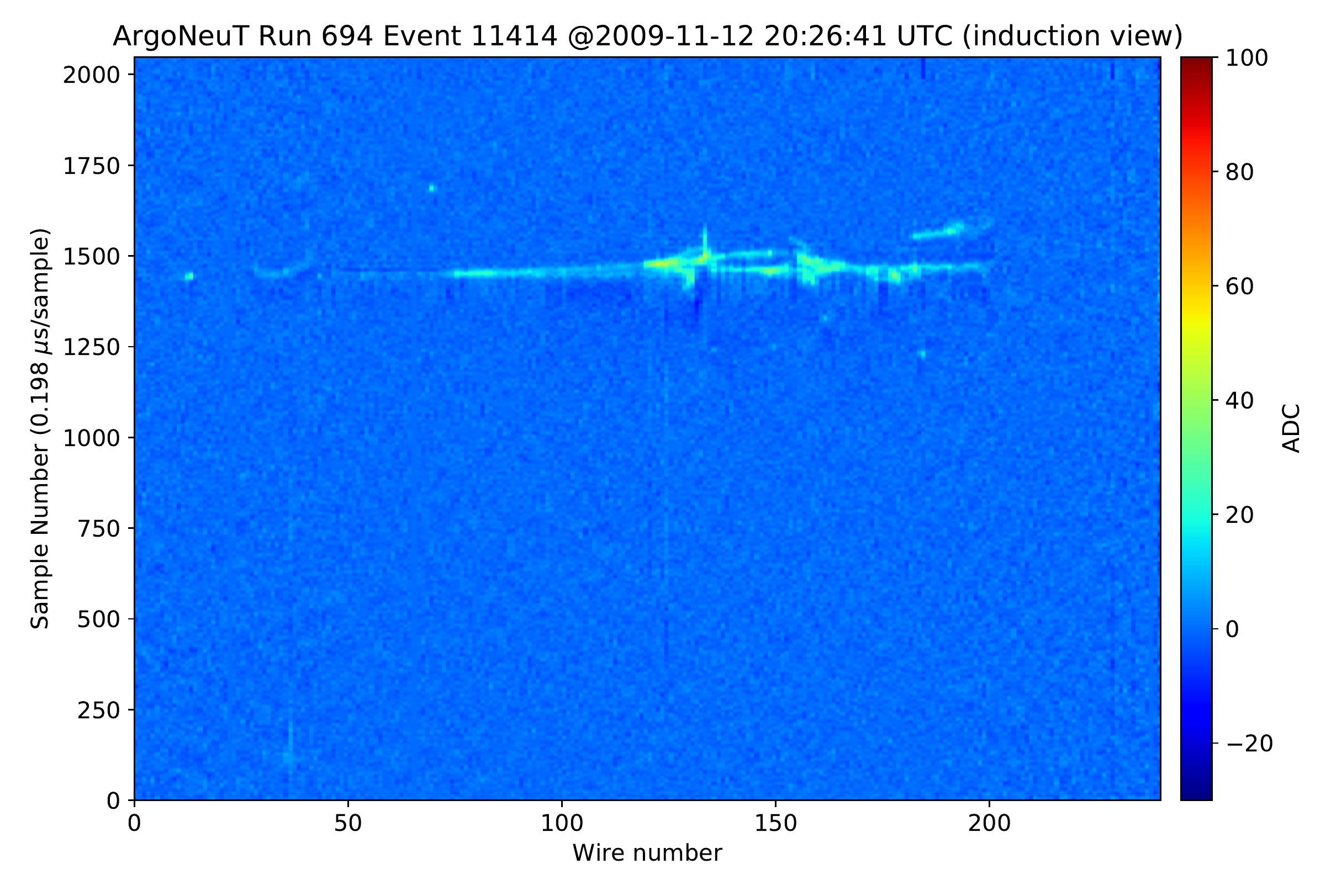} 
\includegraphics[width=.49\textwidth]{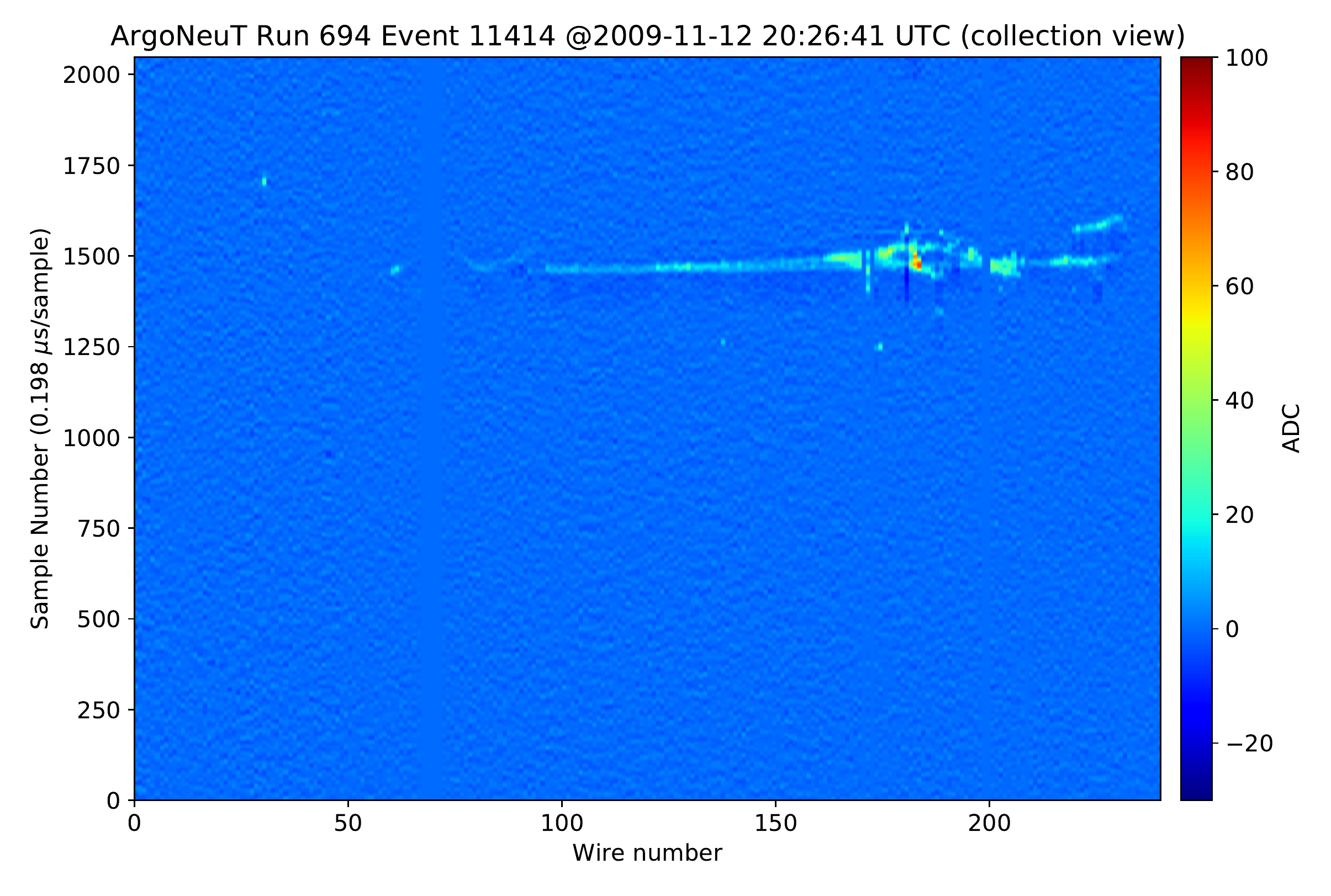} 
\end{figure*}

\begin{figure*}[th!]
\centering 
\includegraphics[width=.49\textwidth]{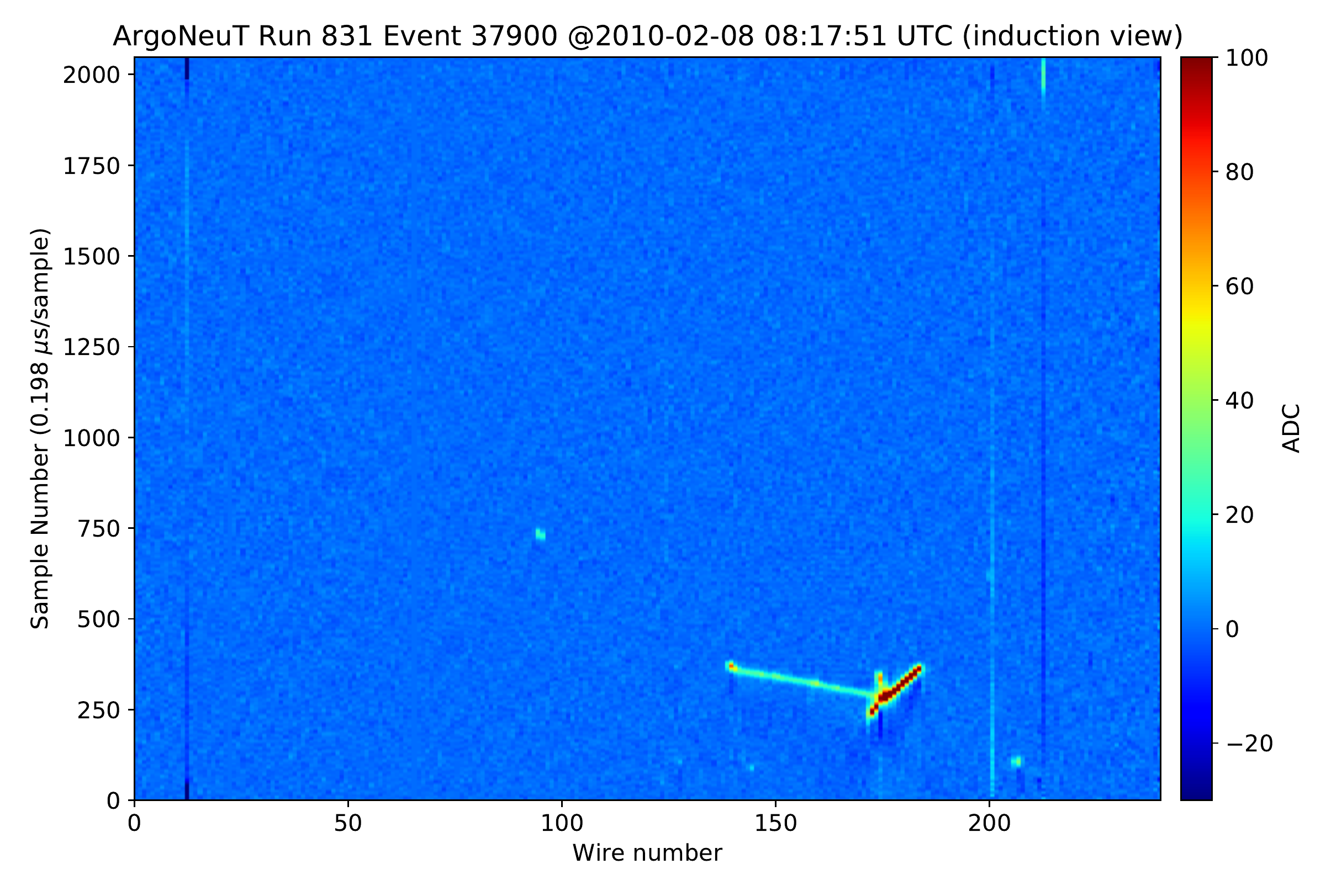} 
\includegraphics[width=.49\textwidth]{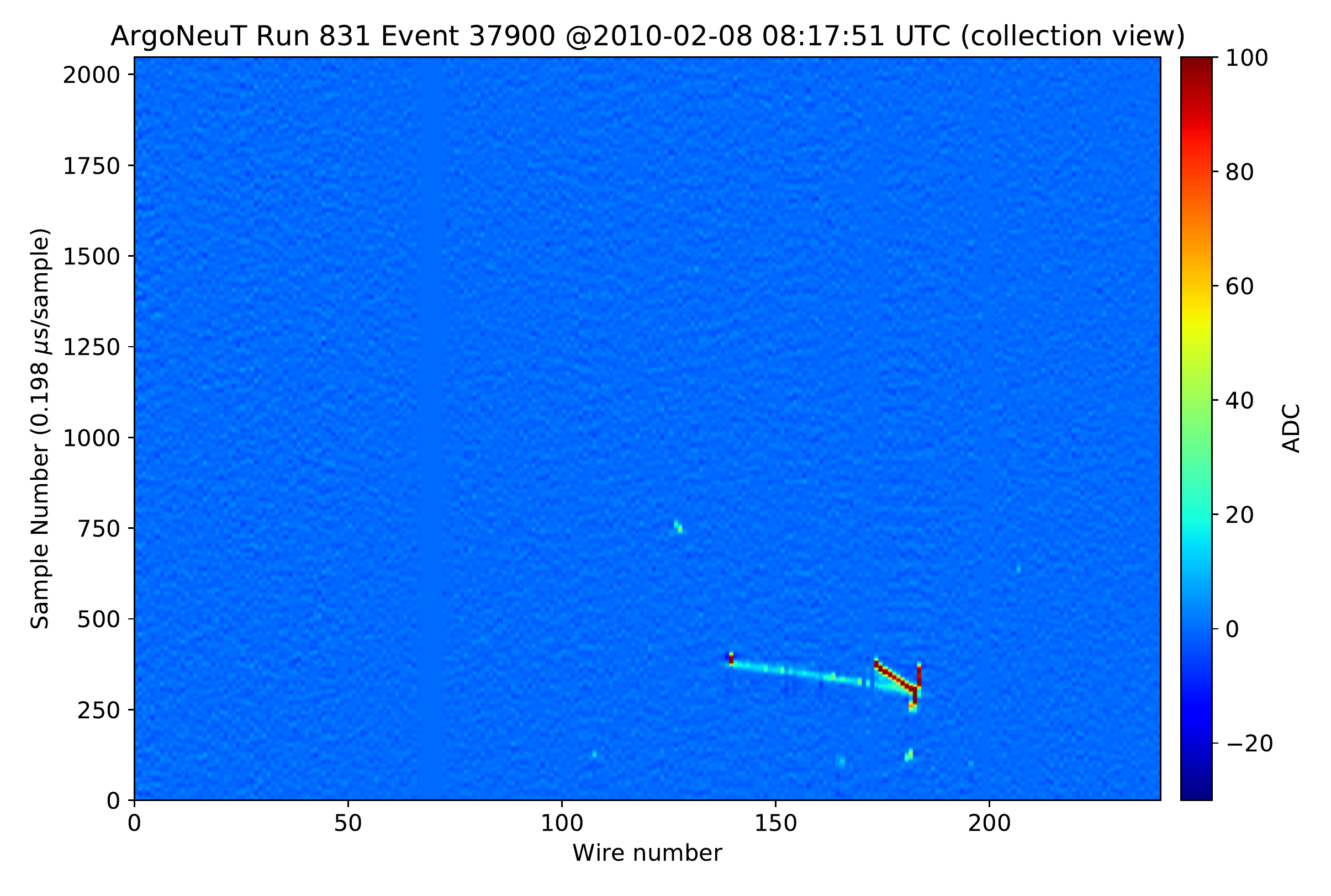} 
\end{figure*}

\end{document}